\documentclass[]{article}

\usepackage{arxiv}
\usepackage{siunitx}
\usepackage{adjustbox}
\usepackage[numbers]{natbib}
\usepackage[utf8]{inputenc} 
\usepackage[T1]{fontenc}    
\usepackage{hyperref}       
\usepackage{url}            
\usepackage{booktabs}       
\usepackage{amsfonts}       
\usepackage{nicefrac}       
\usepackage{microtype}      
\usepackage{lipsum}
\usepackage{fancyhdr}       
\usepackage{graphicx}       
\graphicspath{{media/}}     
\usepackage{enumitem}
\usepackage{multirow}
\usepackage{amsmath}
\usepackage{setspace}
\usepackage[margin=10pt,font=small,labelfont=bf]{caption}
\usepackage{subcaption}
\usepackage[capitalize]{cleveref}
\usepackage[dvipsnames]{xcolor}

\newcommand{\distillation}[1][6pt]{\includegraphics[height=#1]{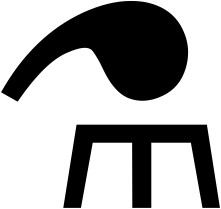}}
\newcommand{\etal}{\textit{et al.}}

\crefname{section}{Sec.}{Secs.}
\crefname{section}{Section}{Sections}
\crefname{table}{Table}{Tables}
\crefname{table}{Tab.}{Tabs.}

\title{Generative Adversarial Super-Resolution at the Edge with Knowledge Distillation}

\author{
  Simone Angarano$^{1,2}$, Francesco Salvetti$^{1,2,3}$, Mauro Martini$^{1,2}$, Marcello Chiaberge$^{1,2}$\\
  $^1$Department of Electronics and Telecommunications, Politecnico di Torino, Turin, Italy\\
  $^2$PIC4SeR PoliTo Interdepartmental Center for Service Robotics\\
  $^3$SmartData@PoliTo, Big Data and Data Science Laboratory\\
  \texttt{\{name.surname\}@polito.it}
}

\begin{document}
\let\WriteBookmarks\relax
\def\floatpagepagefraction{1}
\def\textpagefraction{.001}
\maketitle

\begin{abstract}
Single-Image Super-Resolution can support robotic tasks in environments where a reliable visual stream is required to monitor the mission, handle teleoperation or study relevant visual details. In this work, we propose an efficient Generative Adversarial Network model for real-time Super-Resolution, called EdgeSRGAN\footnote{Code available at \url{https://github.com/PIC4SeR/EdgeSRGAN}.}. We adopt a tailored architecture of the original SRGAN and model quantization to boost the execution on CPU and Edge TPU devices, achieving up to 200 fps inference. We further optimize our model by distilling its knowledge to a smaller version of the network and obtain remarkable improvements compared to the standard training approach. Our experiments show that our fast and lightweight model preserves considerably satisfying image quality compared to heavier state-of-the-art models. Finally, we conduct experiments on image transmission with bandwidth degradation to highlight the advantages of the proposed system for mobile robotic applications.
\end{abstract}


\section{Introduction}\label{sec:intro}
In the last decade, Deep Learning (DL) techniques have pervaded robotic systems and applications, drastically boosting automation in both perception \cite{de2021deep, zhu2021deep}, navigation and control \cite{roy2021survey, xiao2022motion} tasks. The development of Machine Learning driven algorithms is paving the way for advanced levels of autonomy for mobile robots, widely increasing the reliability of both unmanned aerial vehicles (UAV) and unmanned ground vehicles (UGV) \cite{de2021deep}. Nonetheless, the adoption of mobile robots for mapping and exploration \cite{lluvia2021active}, search and rescue \cite{drew2021multi} or inspection \cite{yuan2022novel, yin2021inspection} missions in harsh unseen environments can provide substantial advantages and reduce the risks for human operators. In this context, the successful transmission of images acquired by the robot to the ground station often assumes a significant relevance to the task at hand, allowing the human operators to get real-time information, monitor the state of the mission, take critical planning decisions and analyze the scenario. Moreover, unknown outdoor environments may present unexpected extreme characteristics which still hinder the release of unmanned mobile robots in the complete absence of human supervision. Although novel DL-based autonomous navigation algorithms are currently under investigation in disparate outdoor contexts such as tunnel exploration \cite{rouvcek2019darpa,tardioli2019ground,elmokadem2022method}, row-crops navigation \cite{martini2022position, aghi2021deep} and underwater \cite{li2021underwater,almanza2021deep}, complete or partial remote teleoperation remains the most reliable control strategy in uncertain scenarios. Indeed, irregular terrain, lighting conditions, and the loss of localization signal can lead navigation algorithms to fail. As a direct consequence of navigation errors, the robotic platform can get stuck in critical states where human intervention is required or preferred. 

However, visual data transmission for robot teleoperation, monitoring, or online data processing requires a stable continuous stream of images, which may be drastically affected by poor bandwidth conditions due to the long distance of the robot or by constitutive factors of the specific environment. Besides this, UAVs and high-speed platforms require the pilot to receive the image stream at a high framerate to follow the vehicle's motion in non-line-of-sight situations. A straightforward but effective solution to mitigate poor bandwidth conditions and meet high-frequency transmission requirements is reducing the transmitted image's resolution. On the other hand, heavy image compression with massive loss of detail can compromise image usability. 

\begin{figure}[ht]
	\centering
		\includegraphics[width=0.7\textwidth]{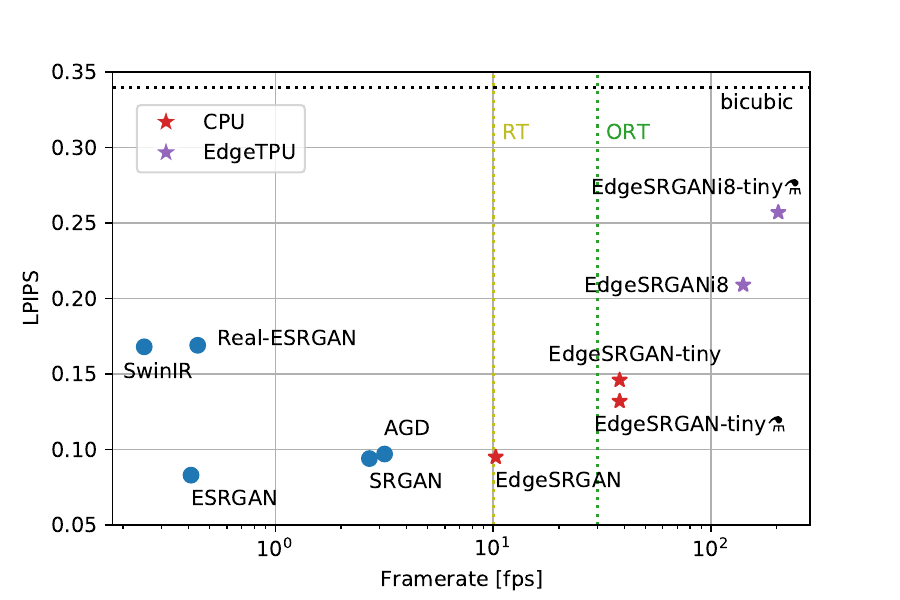}
	  \caption{LPIPS \cite{zhang2018unreasonable} results (lower is better) on Set5 \cite{bevilacqua2012low} vs framerate ($80\times60$ input) of different visual-oriented SISR methods for $\times4$ upsampling. Real-time (RT) and over-real-time (ORT) framerates are marked as references. Our models, marked with $\star$, reach real-time performance with a competitive perceptual similarity index on the CPU. Edge TPU models can further increase inference speed far beyond real-time, still outperforming the bicubic baseline.} \label{fig:fps_lpips}
\end{figure}

To this end, we propose EdgeSRGAN, a novel deep learning model for Single-Image Super-Resolution (SISR) at the edge to handle the problem of efficient image transmission. Our intuition relies on a lightweight neural network allowing us to send low-resolution images at a high transmission rate with scarce bandwidth and then reconstruct the high-resolution image on the pilot's mobile device. Moreover, the successful spread of edge-AI in different engineering applications \cite{chen2019deep, angarano2021robust, liu2021bringing} has shown encouraging results in moving the execution of DL models on ultra-low power embedded devices. Hence, we propose an edge-AI computationally efficient Super Resolution neural network to provide fast inference on CPUs and Edge TPU devices. To this aim, we adopt several optimization steps to boost the performance of our model while minimizing the quality drop. We refine the architecture of the original SRGAN \cite{ledig2017photo} to speed up inference and perform model quantization. Nonetheless, we experiment with a teacher-student knowledge distillation technique for SISR to further enhance the reconstructed image of our tiny model. We take inspiration from the work of \cite{he2020fakd} and obtain a remarkable improvement for all the considered metrics.  

We perform experiments to validate the proposed methodology under multiple perspectives: numerical and qualitative analysis of our model reconstructed images and inference efficiency on both CPU and Edge TPU devices. As an example, as shown in Fig. \ref{fig:fps_lpips}, EdgeSRGAN achieves real-time performance with a competitive perceptual similarity index compared with other visual-oriented SISR methods. Moreover, we test the performance of our system for robotic applications. In particular, we focus on image transmission for teleoperation in case of bandwidth degradation, also performing tests with the popular robotic middleware ROS2.

The rest of the paper is organized as follows. In Section \ref{sec:related-works}, we introduce the research landscape of Super-Resolution (SR), starting from the general background and then deepening the discussion towards robotic applications of SR and efficient SR methods presented in previous works. In Section \ref{sec:methodology}, we describe the Super-Resolution problem and our methodological steps to obtain an Edge AI implementation for real-time performances. In Section \ref{sec:experiments}, we propose a wide range of experiments to validate the proposed methodology, analyzing the results obtained for inference speed and output image quality and characterizing the advantages of our approach for robotic applications in limited-bandwidth conditions. Finally, in Section \ref{sec:concl}, we summarize the overall study with conclusive remarks and suggest some potential future work directions.

\section{Related Works}\label{sec:related-works}

\subsection{Single-Image Super-Resolution}\label{sub-sec:sisr}
Single-Image Super-Resolution, also referred to as super-sampling or image restoration, aims at reconstructing a high-resolution (HR) image starting from a single low-resolution (LR) input image, trying to preserve details and the information conceived by the image. Therefore SISR, together with image denoising, is an ill-posed underdetermined inverse problem since a multiplicity of possible solutions exist given an input low-resolution image. Recently, learning-based methods have rapidly reached state-of-the-art performance and are universally recognized as the most popular approach for Super-Resolution. Such approaches rely on learning common patterns from multiple LR-HR pairs in a supervised fashion. SRCNN \cite{dong2015image} was the first example of a CNN applied to single-image super-resolution in literature. It has been followed by multiple methods applying standard deep learning methodologies to SISR, such as residual learning \cite{kim2016accurate, lim2017enhanced}, dense connections \cite{zhang2018residual}, residual feature distillation \cite{liu2020residual}, attention \cite{zhang2018image, dai2019second, niu2020single}, self-attention, and transformers \cite{cao2021video, chen2021pre, liang2021swinir}. All these works focus on content-based SR, in which the objective is to reconstruct an image with high pixel fidelity, and the training is based on a content loss, such as mean square error or mean absolute error.

In parallel, other works proposed Generative Adversarial Networks (GAN) \cite{goodfellow2014generative} for SISR to aim at reconstructing visually pleasing images. In this case, the focus is not on pixel values but perceptual indexes that try to reflect how humans perceive image quality. This is usually implemented using perceptual losses and adversarial training and is referred to as visual-based SR. SRGAN \cite{ledig2017photo} first proposed adversarial training and was later followed by other works \cite{lim2017enhanced, fu2020autogan, wang2021real-esrgan}. With robotic image transmission as a target application in mind, in this work, we particularly focus on visual-based SR, aiming to reconstruct visually pleasing images to be used by human operators for real-time teleoperation and monitoring.

\subsection{Efficient Methods for Single-Image Super-Resolution}
In recent years, efficient deep neural networks for SR have been proposed to reduce the number of parameters while keeping high-quality performances \cite{li2022ntire}. However, most of the proposed architectural solutions are designed for content-based training, which aims to minimize the difference between the high-resolution image and the network output. Among them, \cite{sajjadi2017enhancenet} proposed a thin, simple model which handles SR as a bilinear upsampling residual compensation. Despite the high-quality images obtained, this approach has high inference latency due to the double prediction required. Diversely, \cite{michelini2022edge} entirely based their study to target Edge-AI chips, proposing an ultra-tiny model composed of one layer only.

As already stated, we prefer GAN-based SR to enhance the visual appearance of produced images for robotic applications. However, successful studies of efficient GANs are very rare in the literature. Recently, knowledge distillation (KD) emerged as a promising option to compress deep models and GANs too \cite{aguinaldo2019compressing,gou2021knowledge}. KD was originally born in 2015 with the visionary work of \cite{hinton2015distilling}, where a teacher-student framework was introduced as a knowledge transfer mechanism. More recent works evolved such concept in disparate variants: FitNets \cite{romero2014fitnets} introduced the idea of involving also intermediate representations in the distillation process, Attention Transfer (AT) \cite{zagoruyko2016paying} proposes an attention-based distillation, and Activation Boundaries (AB) \cite{heo2019knowledge} interestingly focuses on the distilled transfer of activation boundaries formed by hidden neurons, further advanced in \cite{heo2019comprehensive}.
Specifically considering KD application in SR, Feature Affinity KD (FAKD) \cite{he2020fakd} uses intermediate features affinity distillation for PSNR-focused SR. We found this approach a good starting point also for GAN-based SR. Diversely, \cite{zhang2021data} investigates a progressive knowledge distillation method for data-free training. Besides KD, \cite{fu2020autogan} recently proposed an Automated Machine Learning (Auto-ML) framework to search for optimal neural model structure, and filter pruning has been used as another optimization technique \cite{li2016pruning}.

Differently from previous works, our model optimization for edge-SR is composed of three main steps: first, an edge-oriented architectural definition is performed; then, we leverage teacher-student knowledge distillation to further reduce the dimension of our model; lastly, we perform TensorFlow Lite (TFLite) conversion and quantization to shift the network execution to CPUs and Edge TPUs with maximum inference speed.

\subsection{Super-Resolution for Robotic Applications}
SISR has been recently proposed in a few robotic applications where a high level of detail is beneficial to support the specific task.
Research on the indoor teleoperation of mobile robots mainly focuses on improving user experience, combining Deep Learning methods with Virtual Reality \cite{zein2021deep, hedayati2018improving, 8968598}, but neglecting the potential bottleneck caused by connectivity degradation in harsh conditions. Differently, a great effort has been devoted to SISR for underwater robotics perception \cite{ooyama2021underwater, DBLP:journals/corr/abs-2002-01155}, effectively tackling the problem of high-quality image acquisition under the sea for accurate object and species detection.
Besides autonomous navigation applications, interesting contexts are robotic surgery \cite{wang2021real, brodie2018future} and medical robots research \cite{martinez2021super}, where SISR can provide substantial advantages improving the visibility and increasing the level of detail required for delicate high-precision movements of the surgeon. Similarly, a detailed image acquired by a robot is needed for monitoring and inspection purposes. For example, \cite{bae2021deep} uses a Super-Resolution model to enhance the online crack detection and in-situ analysis of bridge weaknesses. Nonetheless, no relevant works proposed so far have identified Super-Resolution as an efficient solution for image transmission to support robot teleoperation and exploration of unknown environments in bandwidth-degraded conditions.

\begin{figure}[t]
	\centering
		\includegraphics[width=0.8\textwidth]{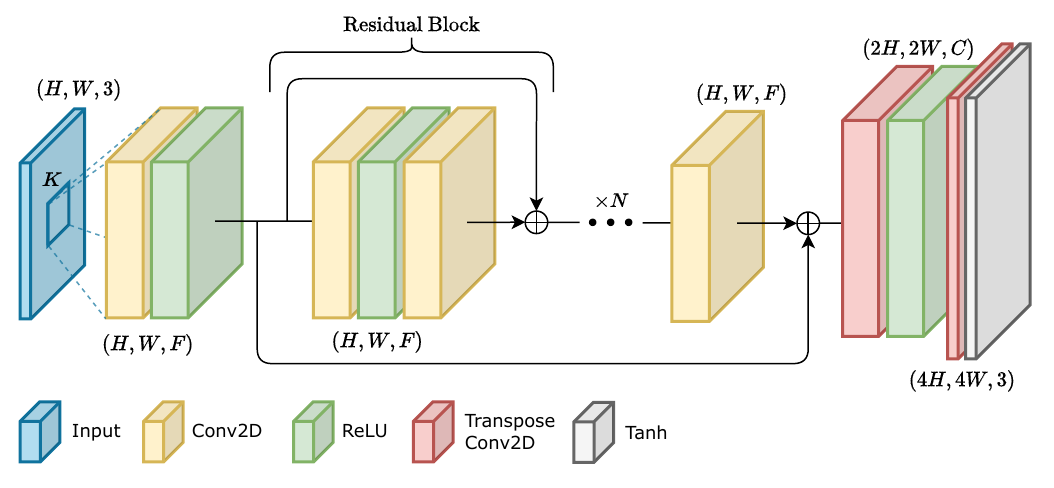}
	  \caption{EdgeSRGAN Generator Architecture.}\label{fig:architecture}
\end{figure}

\section{Methodology}
\label{sec:methodology}
In this section, we introduce all the components of the proposed methodology. As explained in Section \ref{sec:intro}, we choose to use an adversarial approach to obtain an optimal balance between pixel-wise fidelity and perceptual quality. For this reason, we take inspiration from three of the most popular GAN-based solutions for SISR: SRGAN \cite{ledig2017photo}, ESRGAN \cite{wang2018esrgan}, and AGD \cite{fu2020autogan}. The proposed method aims to obtain a real-time SISR model (EdgeSRGAN) with minimal performance drop compared to state-of-the-art solutions. For this reason, we mix successful literature practices with computationally-efficient elements to obtain a lightweight architecture. Then, we design the network training procedure to leverage a combination of pixel-wise loss, perceptual loss, and adversarial loss. To further optimize the inference time, we apply knowledge distillation to transfer the performance of EdgeSRGAN to an even smaller model (EdgeSRGAN-tiny). Furthermore, we study the effect of quantization on the network's latency and accuracy. Finally, we propose an additional inference-time network interpolation feature to allow real-time balancing between pixel-wise precision and photo-realistic textures.

\subsection{Network Architecture} 
\label{sec:architecture}
\begin{figure}[t]
	\centering
		\includegraphics[width=\textwidth]{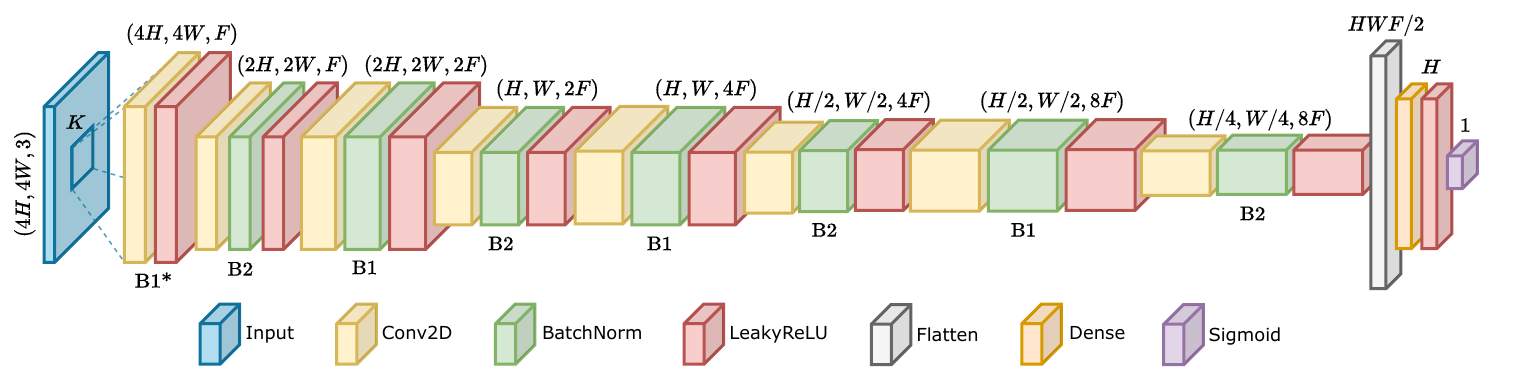}
	  \caption{EdgeSRGAN Discriminator Architecture. The model progressively reduces the spatial dimensions of the image by alternating blocks with strides 1 (B1) and 2 (B2). The first block (marked with *) does not apply batch normalization.}\label{fig:discriminator}
\end{figure}

As previously done by \cite{wang2018esrgan}, we take the original design of SRGAN and propose some changes to both the architecture and training procedure. However, in our case, the modifications seek efficiency as well as performance. To obtain a lighter architecture, we reduce the depth of the model by using only $N=8$ Residual Blocks instead of the original 16. In particular, we use simple residuals instead of the Residual-in-Residual Dense Blocks (RRDB) proposed by \cite{wang2018esrgan} as they are less computationally demanding. For the same reason, we change PReLU activation functions into basic ReLU. We also remove Batch Normalization to allow the model for better convergence without generating artifacts \cite{wang2018esrgan}. Finally, we use Transpose Convolution for the upsampling head instead of Sub-pixel Convolution \cite{Shi_2016_CVPR}. Despite its popularity and effectiveness, Sub-pixel Convolution is computationally demanding due to the Pixel Shuffling operation, which rearranges feature channels spatially. We choose instead to trade some performance for efficiency and apply Transpose Convolutions taking precautions to avoid problems such as checkerboard artifacts \cite{odena2016deconvolution}. The complete EdgeSRGAN architecture is described in Fig. \ref{fig:architecture}. The adopted discriminator model is the same used in \cite{ledig2017photo, wang2018esrgan}, as it serves only training purposes and is not needed at inference time. Its architecture is described in Fig. \ref{fig:discriminator}.

\subsection{Training Methodology} 
\label{sec:training}

The training procedure is divided into two sections, as it is common practice in generative adversarial SISR. The first part consists of classic supervised training using pixel-wise loss. In this way, we help the generator to avoid local minima and generate visually pleasing results in the subsequent adversarial training. We use the mean absolute error (MAE) loss for the optimization as it has recently proven to bring better convergence than mean squared error (MSE) \cite{zhao2016loss, lim2017enhanced, zhang2018image, wang2018esrgan}.
\begin{equation}
    L_{\text{MSE}} = \sum_{i=1}^{B}|| y^{\text{HR}}_i - y^{\text{SR}}_i ||_1
\end{equation}
where $y^\text{HR}$ is the ground-truth high resolution image, $y^{\text{SR}}$ is the output of the generator, and $B$ is the batch size. We use the Peak Signal-to-Noise Ratio (PSNR) metric to validate the model.

\begin{figure}[t]
	\centering
		\includegraphics[width=\textwidth]{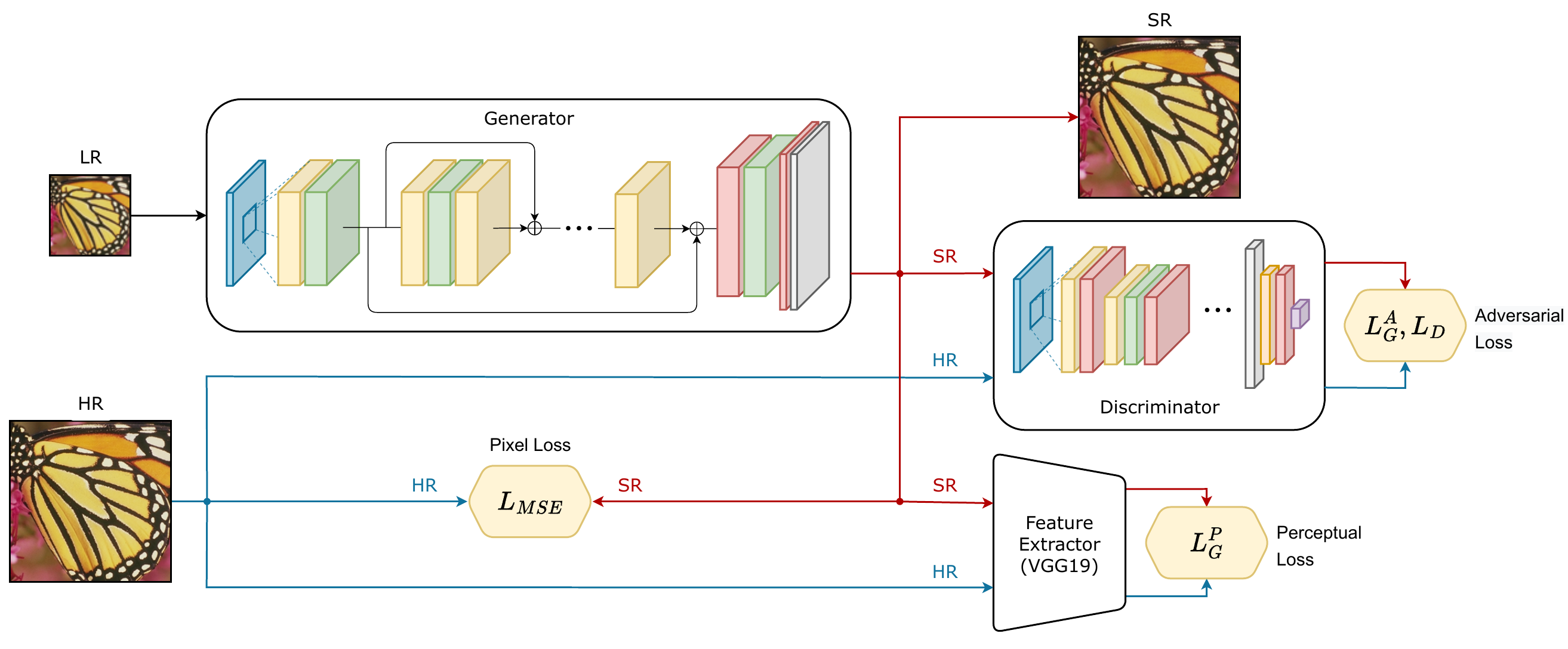}
	  \caption{EdgeSRGAN Training Methodology.}\label{fig:allinone}
\end{figure}

In the second phase, the resulting model is fine-tuned in an adversarial fashion, optimizing a loss that takes into account adversarial loss and perceptual loss. As presented in \cite{ledig2017photo}, the generator $G$ training loss can be formulated as
\begin{equation}
    L_G = L_G^P + \xi L_G^A + \eta L_{\text{MSE}}.
\end{equation}
$L_G^P$ is the perceptual VGG54  as the euclidean distance between the feature representations of a reconstructed image SR and the reference image HR. The features are extracted using the VGG19 network \cite{vgg2015} pre-trained on ImageNet:
\begin{equation}
    L_G^P = \sum_{i=1}^{B}|| \phi(y^{\text{HR}}_i) - \phi(y^{\text{SR}}_i) ||_2
\end{equation}
\noindent where $\phi$ is the perceptual model VGG. $L_G^A$ is the adversarial generator loss, defined as
\begin{equation}
    L_G^A = - \log(D(y_{\text{SR}}))
\end{equation}
\noindent where $D$ is the discriminator. Using this loss, the generator tries to fool the discriminator by generating images that are indistinguishable from the real HR ones. $\xi$ and $\eta$ are used to balance the weight of different loss components.
The weights of the discriminator $D$ are optimized using a symmetrical adversarial loss, which tends to correctly discriminate HR and SR images.
\begin{equation}
    L_D = \log(D(y_{\text{SR}})) - \log(D(y_{\text{HR}}))
\end{equation}
We optimize both models simultaneously, without alternating weight updates like in most seminal works on GANs. The overall training methodology is summarized in Fig. \ref{fig:allinone} summarizes the overall training methodology.

\subsection{Knowledge Distillation}
\label{sec:distillation}

\begin{figure}[t]
	\centering
		\includegraphics[width=\textwidth]{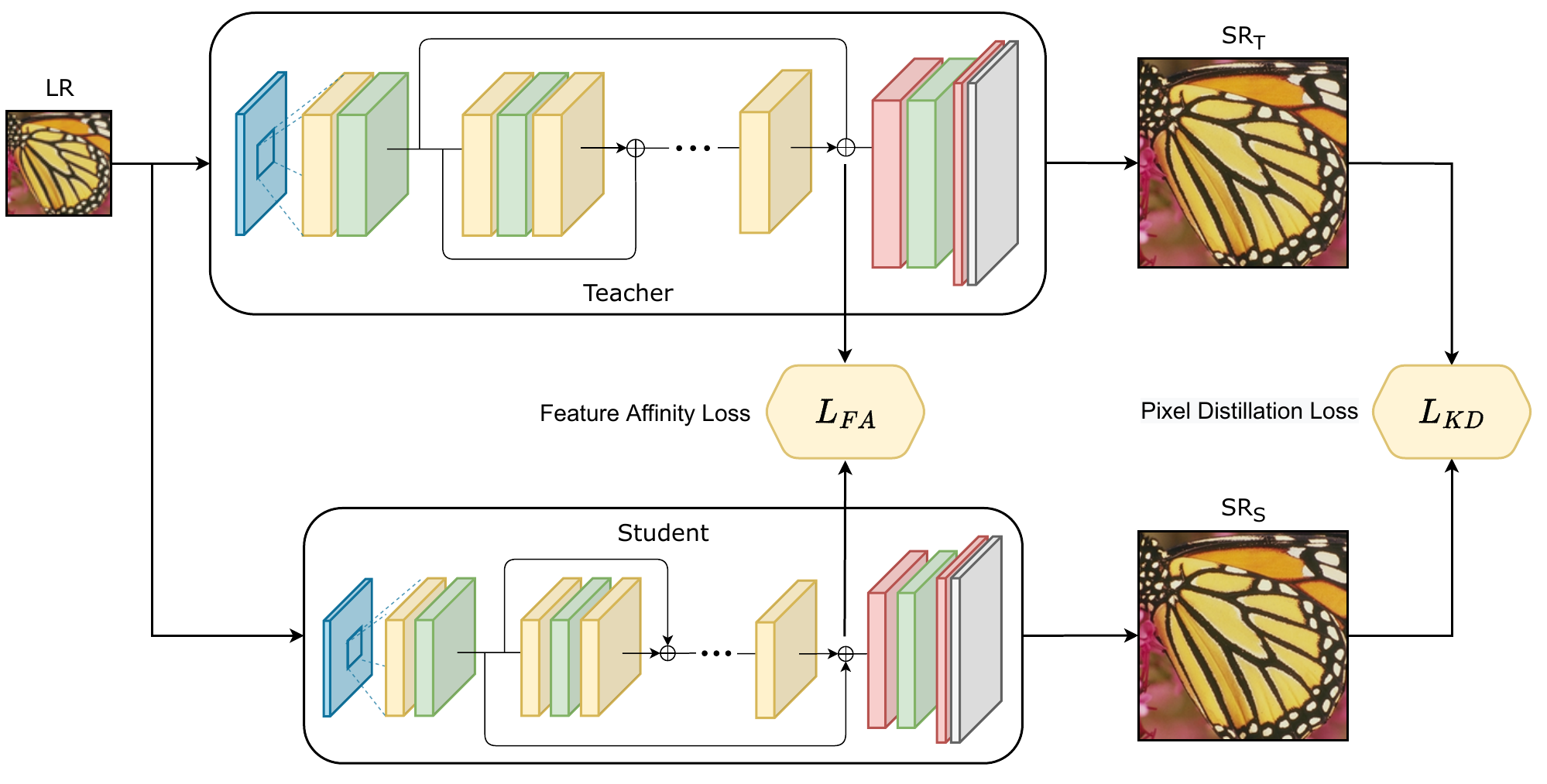}
	  \caption{EdgeSRGAN Distillation Process.}\label{fig:dist}
\end{figure}

As mentioned in Section \ref{sec:related-works}, Knowledge Distillation (KD) has gained increasing interest in deep learning for its ability to transfer knowledge from bigger models to simpler ones efficiently. In particular, KD has been applied in some SISR works to compress the texture reconstruction capability of cumbersome models and obtain efficient real-time networks. However, to the best of our knowledge, KD has never been applied to GAN SISR models. For this reason, we adapt an existing technique developed for SISR called Feature Affinity-based Knowledge Distillation (FAKD) \cite{he2020fakd} to the GAN training approach. The FAKD methodology transfers second-order statistical info to the student by aligning feature affinity matrices at different layers of the networks. This constraint helps to tackle the fact that regression problems generate unbounded solution spaces. Indeed, most of the KD methods so far have only tackled classification tasks. 
Given a layer $l$ of the network, the feature map $F_l$ extracted from that layer (after the activation function) has the following shape:
\begin{equation}
    F_l \in \mathbb{R}^{B\times C\times W\times H}
\end{equation}
\noindent where $B$ is the batch size, $C$ is the number of channels, $W$ and $H$ are the width and the height of the tensor. We first flatten the tensor along the last two components obtaining the three-dimensional feature map
\begin{equation}
    F_l \in \mathbb{R}^{B\times C\times \textit{WH}}
\end{equation}
\noindent which now holds all the spatial information along a single axis. We define the affinity matrix $A_l$ as the product
\begin{equation}
    A_l = \Tilde{F_l}^\top \cdot \Tilde{F_l}
\end{equation}
\noindent where $\cdot$ is the matrix multiplication operator and the transposition $\top$ swaps the last two dimensions of the tensor. $\Tilde{F_l}$ is the normalized feature map, obtained as 
\begin{equation}
    \Tilde{F_l}=\frac{F_l}{||F_l||_2}
\end{equation}
\noindent Differently from \cite{he2020fakd}, the norm is calculated for the whole tensor and not only along the channel axis. Moreover, we find better convergence using the euclidean norm instead of its square. In this way, the affinity matrix has a shape 
\begin{equation}
A_l \in \mathbb{R}^{B\times \textit{WH}\times \textit{WH}}
\end{equation}
and the total distillation loss $L_{\text{Dist}}$ becomes
\begin{equation}
\label{eq:distil_loss}
    L_{\text{Dist}} = \frac{1}{N_L} \Biggl(\sum_{l=1}^{N_L} ||A^T_l - A^S_l||_1\Biggr) + \lambda ||y^T_\text{SR} - y^S_\text{SR}||_1
\end{equation}
\noindent where $N_L$ is the number of distilled layers. Differently from \cite{he2020fakd}, we sum the loss along all the tensor dimensions and average the result obtained for different layers. These modifications experimentally lead to better training convergence. We also add another loss component, weighted by $\lambda$, which optimizes the model to generate outputs close to the teacher's ones. In our experimentation, the distillation loss is added to the overall training loss weighted by the parameter $\gamma$.  The overall distillation scheme is summarized in Fig. \ref{fig:dist}.

\subsection{Model Interpolation} 
\label{sec:interpolation}
Following the procedure proposed in \cite{wang2018esrgan}, we adopt a flexible and effective strategy to obtain a tunable trade-off between a content-oriented and GAN-trained model. This feature can be very useful for real-time applications, as it allows the SISR network to adapt to the user's needs promptly. Indeed, some real scenarios may need better perceptual quality, for example, when the remote control of a robot has to be performed by a human pilot. On the other hand, when images are used to directly feed perception, autonomous navigation, and mapping algorithms, higher pixel fidelity might be beneficial. To achieve this goal, we linearly interpolate model weights layer-by-layer, according to the following formula:
\begin{equation}
    \theta^{\text{Int}}_G = \alpha\theta^{\text{PSNR}}_G + (1-\alpha) \theta^{\text{GAN}}_G
\end{equation}
\noindent where $\theta^{\text{Interp}}_G$, $\theta^{\text{PSNR}}_G$, and $\theta^{\text{GAN}}_G$ are the weights of the interpolated model, the PSNR model, and the GAN fine-tuned model, respectively. $\alpha \in [0,1]$ is the interpolation weight. We report both qualitative and quantitative interpolation results for EdgeSRGAN in Section \ref{sec:NIresults}. We avoid the alternative technique of directly interpolating network outputs: applying this method in real time would require running two models simultaneously. Moreover, Wang \etal \cite{wang2018esrgan} report that this approach does not guarantee an optimal trade-off between noise and blur.

\subsection{Model Quantization} 
\label{secf:optimization}
To make EdgeSRGAN achieve even lower inference latency, we apply optimization methods to the model to reduce the computational effort at the cost of a loss in performance. Several techniques have been developed to increase model efficiency in the past few years \cite{jacob2018quantization}, from which the employed method is chosen. We reduce the number of bits used to represent network parameters and activation functions with TFLite\footnote{\url{https://www.tensorflow.org/lite/}}. This strategy strongly increases efficiency with some impact on performance. 
We quantize weights, activations, and math operations through scale and zero-point parameters following the methodology presented by Jacob \etal \cite{jacob2018quantization}:
\begin{equation}
\label{eq:quantization}
    r = S(q - Z)
\end{equation}
where $r$ is the original floating-point value, $q$ is the quantized integer value, and $S$ and $Z$ are the quantization parameters (scale and zero point).
A fixed-point multiplication approach is adopted to cope with the non-integer scale of $S$. This strategy drastically reduces memory and computational demands due to the high efficiency of integer computations on microcontrollers. For our experimentation, we deploy the quantized model on a Google Coral Edge TPU USB Accelerator\footnote{\url{https://coral.ai/}}. 

\begin{table}[ht]
\centering
\resizebox{0.7\textwidth}{!}{
\begin{tabular}{@{}lcccccc@{}}
&  &  & \multicolumn{2}{c}{\textbf{Framerate ($80\times60$)} [fps]} & \multicolumn{2}{c}{\textbf{Framerate ($160\times120$)} [fps]} \\ \cmidrule(l){4-7} 
\textbf{Method}  & \textbf{Scale}  & \textbf{Params}       & \textbf{CPU}     & \textbf{EdgeTPU}     & \textbf{CPU}      & \textbf{EdgeTPU}      \\ \midrule
\textbf{SwinIR \cite{liang2021swinir}} & \multicolumn{1}{|c|}{\multirow{7}{*}{$\times4$}} & 11.9M  & 0.25 ± 0.01  & -  & 0.06 ± 0.01  & -   \\
\textbf{ESRGAN \cite{wang2018esrgan}}  & \multicolumn{1}{|c|}{} & 16.7M & 0.40 ± 0.01  & -  & 0.10 ± 0.01  & -   \\
\textbf{Real-ESRGAN \cite{wang2021real-esrgan}}  & \multicolumn{1}{|c|}{} & 16.7M & 0.44 ± 0.01  & -  & 0.11 ± 0.01  & -   \\
\textbf{SRGAN \cite{ledig2017photo}} & \multicolumn{1}{|c|}{} & 1.5M   & 2.70 ± 0.08  & -  & 0.95 ± 0.02  & -   \\
\textbf{AGD \cite{fu2020autogan}}    & \multicolumn{1}{|c|}{} & 0.42M  & 3.17 ± 0.12  & -  & 0.88 ± 0.01  & -   \\
\textbf{EdgeSRGAN}  & \multicolumn{1}{|c|}{} & 0.66M & \textcolor{blue}{10.26 ± 0.11} & \textcolor{red}{140.23 ± 1.50}  & 2.66 ± 0.02  & \textcolor{blue}{10.63 ± 0.03}   \\
\textbf{EdgeSRGAN-tiny} & \multicolumn{1}{|c|}{} & 0.09M & \textcolor{red}{37.99 ± 1.42}   & \textcolor{red}{203.16 ± 3.03}  & \textcolor{blue}{11.76 ± 0.20} & \textcolor{blue}{20.57 ± 0.05}   \\ \midrule \midrule

\textbf{SwinIR \cite{liang2021swinir}} & \multicolumn{1}{|c|}{\multirow{3}{*}{$\times8$}} & 12.0M & 0.23 ± 0.01  & -  & 0.06 ± 0.01  & -   \\
\textbf{EdgeSRGAN} & \multicolumn{1}{|c|}{} & 0.71M & 7.70 ± 0.31  & \textcolor{blue}{14.26 ± 0.06}  & 1.81 ± 0.04  & -   \\
\textbf{EdgeSRGAN-tiny} & \multicolumn{1}{|c|}{} & 0.11M & \textcolor{blue}{24.53 ± 1.28}  & \textcolor{red}{41.55 ± 0.38}  & 5.81 ± 0.29 & -   \\ \bottomrule
\end{tabular}
}
\caption{Framerate comparison of different methods for $\times 4$ and $\times 8$ upsampling, with two different input resolutions ($80\times60$ and $160\times120$). The results are provided as mean and standard deviation of 10 independent experiments of 100 predictions each. Current content-oriented SISR state-of-art method SwinIR \cite{liang2021swinir} is reported as a reference. Real-time and over-real-time framerates are in \textcolor{blue}{blue} and \textcolor{red}{red}, respectively. The proposed solution is the only one compatible with EdgeTPU devices and allows reaching real-time performance in both conditions.}
\label{tab:fps}
\end{table}

\section{Experiments}
\label{sec:experiments}

\subsection{Experimental Setting}
\label{sec:setting}
In this section, we define our method's implementation details and the procedure we followed to train and validate the efficiency of EdgeSRGAN optimally. As previously done by most GAN-based SISR works, we train the network on the high-quality DIV2K dataset \cite{agustsson2017ntire} with a scaling factor of 4. The dataset contains 800 training samples and 100 validation samples. We train our model with input images of size 24x24 pixels, selecting random patches from the training set. We apply data augmentation by randomly flipping or rotating the images by multiples of $90^{\circ}$. We adopt a batch size of 16.

For the standard EdgeSRGAN implementation, we choose $N=8$, $F=64$, $K=3$, and $D=1024$, obtaining a generator with around 660k parameters and a discriminator of over 23M (due to the fully-connected head). The discriminator is built with $F=64$, $K=3$, $D=512$, and with a coefficient for LeakyReLU $\alpha=0.2$.
We first train EdgeSRGAN pixel-wise for \num{5e5} steps with Adam optimizer and a constant learning rate of \num{1e-4}. Then, the model is fine-tuned in the adversarial setting described in Section \ref{sec:methodology} for \num{1e5} steps. Adam optimizer is used for the generator and the discriminator with a learning rate of \num{1e-5}, further divided by 10 after \num{5e4} steps. For the loss function, we set $\xi=\num{1e-3}$ and $\eta=0$.

To obtain an even smaller model for our distillation experiments, we build EdgeSRGAN-tiny by choosing $N=4$, $F=32$, and $D=256$. We further shrink the size of the discriminator by eliminating the first compression stage ($B1$) from each block (see Fig. \ref{fig:discriminator}). In this configuration, we also remove the batch normalization layer from the first B2 block to be coherent with the larger version. The obtained generator and discriminator contain around 90k and 2.75M parameters. The pre-training procedure is the one described for EdgeSRGAN, while the adversarial training is performed with the additional distillation loss ($\gamma=\num{1e-2}$, $\lambda=\num{1e-1}$) of Eq. \ref{eq:distil_loss}. EdgeSRGAN is used as a teacher model, distilling its layers 2, 5, and 8 into EdgeSRGAN-tiny's layers 1, 2, and 4. The model is trained with a learning rate of \num{1e-4}, which is further divided by 10 after \num{5e4} steps. For the loss function, we set $\xi=\num{1e-3}$ and $\eta=0$.

Finally, we create a third version of our model to upscale images with a factor of 8. To do so, we change the first transpose convolution layer of EdgeSRGAN and EdgeSRGAN-tiny to have a stride of 4 instead of 2 and leave the rest of the architecture unchanged. The training procedure for these models is analogous to the ones used for the x4 models, with the main difference of adding a pixel-based component to the adversarial loss by posing $\eta=\num{1e2}$.

The optimal training hyperparameters are found by running a random search and choosing the best-performing models on DIV2K validation. During GAN training, we use PSNR to validate the models during content-based loss optimization and LPIPS \cite{zhang2018unreasonable} (with AlexNet backbone). 

We employ TensorFlow 2 and a workstation with 64 GB of RAM, an Intel i9-12900K CPU, and an Nvidia 3090 RTX GPU to perform all the training experiments.

\subsection{Real-time Performance} 
\label{sec:RTresults}
Since the main focus of the proposed methodology is to train an optimized SISR model to be efficiently run at the edge in real time, we first report an inference speed comparison between the proposed method and other literature methodologies. All the results are shown in Tab. \ref{tab:fps} as the mean and standard deviation of 10 independent experiments of 100 predictions each. We compare the proposed methodology with other GAN-based methods \cite{ledig2017photo,wang2018esrgan,wang2021real-esrgan,fu2020autogan} and with the current state-of-the-art in content-oriented SISR SwinIR \cite{liang2021swinir}. Since the original implementations of the GAN-based solutions consider $\times 4$ upsampling only, for the $\times 8$ comparison, we only report SwinIR. We select two different input resolutions for the experimentation, $(80\times60)$ and $(160\times120)$, in order to target $(320\times240)$ and $(640\times480)$ resolutions for $\times 4$ upsampling and $(640\times480)$ and $(1280\times960)$ for $\times 8$ upsampling, respectively. This choice is justified because $(640\times480)$ is a standard resolution provided by most cameras' native video stream. We also report the number of parameters for all the models.

For all the considered methods, we measure the CPU timings with the model format of the original implementation (PyTorch or TensorFlow) on a MacBook Pro with an Intel i5-8257U processor. The concept of real-time performance strongly depends on the downstream task. For robotic monitoring and teleoperation, we consider 10 fps as the minimum real-time framerate, considering over-real-time everything above 30 fps, which is the standard framerate for most commercial cameras. The proposed methodology outperforms all the other methods in inference speed and achieves real-time performance on the CPU in almost all the testing conditions. It is worth noting that AGD is specifically designed to reduce latency for GAN-based SR and has fewer parameters than EdgeSRGAN, but it still fails at achieving real-time without a GPU.

In addition, we report the framerate of the EdgeSRGAN int8-quantized models on an EdgeTPU Coral USB Accelerator. The proposed solution is the only one compatible with such devices and allows reaching over-real-time performance for $(80\times60)$ input resolution. It must be underlined how the $\times 8$ models with $(160\times120)$ input resolution cannot target the EdgeTPU device due to memory limitations.

\begin{table}[t]
\resizebox{\textwidth}{!}{
\begin{tabular}{@{}lccccccccccccccc@{}}
& \multicolumn{3}{c}{\textbf{Set5} \cite{bevilacqua2012low}}  & \multicolumn{3}{c}{\textbf{Set14} \cite{zeyde2010single}} & \multicolumn{3}{c}{\textbf{BSD100} \cite{martin2001database}}  & \multicolumn{3}{c}{\textbf{Manga109} \cite{matsui2017sketch}} & \multicolumn{3}{c}{\textbf{Urban100} \cite{huang2015single}} \\ \midrule

\multicolumn{1}{l|}{\textbf{Method}}                  & \textbf{PSNR ↑} & \textbf{SSIM ↑} & \multicolumn{1}{c|}{\textbf{LPIPS ↓}} & \textbf{PSNR ↑} & \textbf{SSIM ↑} & \multicolumn{1}{c|}{\textbf{LPIPS ↓}} & \textbf{PSNR ↑} & \textbf{SSIM ↑} & \multicolumn{1}{c|}{\textbf{LPIPS ↓}} & \textbf{PSNR ↑} & \textbf{SSIM ↑} & \multicolumn{1}{c|}{\textbf{LPIPS ↓}} & \textbf{PSNR ↑} & \textbf{SSIM ↑} & \textbf{LPIPS ↓} \\ \midrule

\multicolumn{1}{l|}{\textbf{Bicubic}}                & 28.632          & 0.814           & \multicolumn{1}{c|}{0.340}            & 26.212          & 0.709           & \multicolumn{1}{c|}{0.441}            & 26.043          & 0.672           & \multicolumn{1}{c|}{0.529}            & 25.071          & 0.790           & \multicolumn{1}{c|}{0.318}            & 23.236          & 0.661           & 0.473            \\ 
\multicolumn{1}{l|}{\textbf{SwinIR} \cite{liang2021swinir}}                 & 32.719          & 0.902           & \multicolumn{1}{c|}{0.168}            & 28.939          & 0.791           & \multicolumn{1}{c|}{0.268}            & 27.834          & 0.746           & \multicolumn{1}{c|}{0.358}            & 31.678          & 0.923           & \multicolumn{1}{c|}{0.094}            & 27.072          & 0.816           & 0.193            \\ \midrule

\multicolumn{1}{l|}{\textbf{SRGAN}  \cite{ledig2017photo}}  & 32.013               & 0.893                & \multicolumn{1}{c|}{0.191} & 28.534               & 0.781                & \multicolumn{1}{c|}{0.294} & 27.534               & 0.735                & \multicolumn{1}{c|}{0.396} & 30.292               & 0.906                & \multicolumn{1}{c|}{0.111} & 25.959               & 0.782                & 0.244                \\
\multicolumn{1}{l|}{\textbf{ESRGAN} \cite{wang2018esrgan}$\dag$}  & 32.730               & 0.901                & \multicolumn{1}{c|}{0.181} & 28.997               & 0.792                & \multicolumn{1}{c|}{0.275} & 27.838               & 0.745                & \multicolumn{1}{c|}{0.371} & 31.644               & 0.920                & \multicolumn{1}{c|}{0.097} & 27.028               & 0.815                & 0.201                \\
\multicolumn{1}{l|}{\textbf{AGD} \cite{fu2020autogan}} & 31.708               & 0.889                & \multicolumn{1}{c|}{0.178} & 28.311               & 0.775                & \multicolumn{1}{c|}{0.291} & 27.374               & 0.729                & \multicolumn{1}{c|}{0.385} & 29.413               & 0.897                & \multicolumn{1}{c|}{0.118} & 25.506               & 0.767                & 0.250                \\\midrule

\multicolumn{1}{l|}{\textbf{EdgeSRGAN}} & 31.729               & 0.889                & \multicolumn{1}{c|}{0.191} & 28.303               & 0.774                & \multicolumn{1}{c|}{0.301} & 27.359               & 0.728                & \multicolumn{1}{c|}{0.405} & 29.611               & 0.897                & \multicolumn{1}{c|}{0.120} & 25.469               & 0.764                & 0.266                \\
\multicolumn{1}{l|}{\textbf{EdgeSRGAN-tiny}} & 30.875               & 0.873                & \multicolumn{1}{c|}{0.204} & 27.796               & 0.761                & \multicolumn{1}{c|}{0.320} & 26.999               & 0.717                & \multicolumn{1}{c|}{0.418} & 28.233               & 0.871                & \multicolumn{1}{c|}{0.163} & 24.695               & 0.733                & 0.325                \\ \bottomrule
\end{tabular}
}
\caption{Quantitative comparison of different methods for content-oriented $\times 4$ upsampling. Current SISR state-of-art method SwinIR \cite{liang2021swinir} and bicubic baseline are reported as reference. \\ \scalebox{0.8}{\textbf{↑}}: higher is better, \scalebox{0.8}{\textbf{↓}}: lower is better, $\dag$: trained on DIV2K \cite{agustsson2017ntire} + Flickr2K \cite{timofte2017ntire} + OST \cite{wang2018recovering}}
\label{tab:SRresPSNR}
\end{table}

\begin{table}[t]
\resizebox{\textwidth}{!}{
\begin{tabular}{@{}lccccccccccccccc@{}}
& \multicolumn{3}{c}{\textbf{Set5} \cite{bevilacqua2012low}}  & \multicolumn{3}{c}{\textbf{Set14} \cite{zeyde2010single}} & \multicolumn{3}{c}{\textbf{BSD100} \cite{martin2001database}}  & \multicolumn{3}{c}{\textbf{Manga109} \cite{matsui2017sketch}} & \multicolumn{3}{c}{\textbf{Urban100} \cite{huang2015single}} \\ \midrule

\multicolumn{1}{l|}{\textbf{Model}}                  & \textbf{PSNR ↑} & \textbf{SSIM ↑} & \multicolumn{1}{c|}{\textbf{LPIPS ↓}} & \textbf{PSNR ↑} & \textbf{SSIM ↑} & \multicolumn{1}{c|}{\textbf{LPIPS ↓}} & \textbf{PSNR ↑} & \textbf{SSIM ↑} & \multicolumn{1}{c|}{\textbf{LPIPS ↓}} & \textbf{PSNR ↑} & \textbf{SSIM ↑} & \multicolumn{1}{c|}{\textbf{LPIPS ↓}} & \textbf{PSNR ↑} & \textbf{SSIM ↑} & \textbf{LPIPS ↓} \\ \midrule
\multicolumn{1}{l|}{\textbf{Bicubic}}                & 28.632          & 0.814           & \multicolumn{1}{c|}{0.340}            & 26.212          & 0.709           & \multicolumn{1}{c|}{0.441}            & 26.043          & 0.672           & \multicolumn{1}{c|}{0.529}            & 25.071          & 0.790           & \multicolumn{1}{c|}{0.318}            & 23.236          & 0.661           & 0.473            \\ 
\multicolumn{1}{l|}{\textbf{SwinIR} \cite{liang2021swinir}}                 & 32.719          & 0.902           & \multicolumn{1}{c|}{0.168}            & 28.939          & 0.791           & \multicolumn{1}{c|}{0.268}            & 27.834          & 0.746           & \multicolumn{1}{c|}{0.358}            & 31.678          & 0.923           & \multicolumn{1}{c|}{0.094}            & 27.072          & 0.816           & 0.193            \\ \midrule
\multicolumn{1}{l|}{\textbf{SRGAN}  \cite{ledig2017photo}}                  & 29.182          & 0.842           & \multicolumn{1}{c|}{0.094}            & 26.171          & 0.701           & \multicolumn{1}{c|}{0.172}            & 25.447          & 0.648           & \multicolumn{1}{c|}{0.206}            & 27.346          & 0.860           & \multicolumn{1}{c|}{0.076}            & 24.393          & 0.728           & 0.158            \\
\multicolumn{1}{l|}{\textbf{ESRGAN}\cite{wang2018esrgan}$\dag$}                 & 30.459          & 0.852           & \multicolumn{1}{c|}{0.083}            & 26.283          & 0.698           & \multicolumn{1}{c|}{0.139}            & 25.288          & 0.649           & \multicolumn{1}{c|}{0.168}            & 28.478          & 0.860           & \multicolumn{1}{c|}{0.065}            & 24.350          & 0.733           & 0.125            \\
\multicolumn{1}{l|}{\textbf{Real-ESRGAN} \cite{wang2021real-esrgan}$\dag$}   & 26.617               & 0.807                & \multicolumn{1}{c|}{0.169} & 25.421               & 0.696                & \multicolumn{1}{c|}{0.234} & 25.089               & 0.653                & \multicolumn{1}{c|}{0.282} & 25.985               & 0.836                & \multicolumn{1}{c|}{0.149} & 22.671               & 0.686                & 0.214                \\
\multicolumn{1}{l|}{\textbf{AGD} \cite{fu2020autogan}}                    & 30.432          & 0.861           & \multicolumn{1}{c|}{0.097}            & 27.276          & 0.739           & \multicolumn{1}{c|}{0.160}            & 26.219          & 0.688           & \multicolumn{1}{c|}{0.214}            & 28.163          & 0.870           & \multicolumn{1}{c|}{0.076}            & 24.732          & 0.743           & 0.170            \\  \midrule

\multicolumn{1}{l|}{\textbf{EdgeSRGAN}} & 29.487          & 0.837           & \multicolumn{1}{c|}{0.095}            & 26.814          & 0.715           & \multicolumn{1}{c|}{0.176}            & 25.543          & 0.644           & \multicolumn{1}{c|}{0.210}            & 27.679          & 0.855           & \multicolumn{1}{c|}{0.081}            & 24.268          & 0.716           & 0.170            \\
\multicolumn{1}{l|}{\textbf{EdgeSRGAN-tiny}}  & 28.074          & 0.803           & \multicolumn{1}{c|}{0.146}            & 26.001          & 0.702           & \multicolumn{1}{c|}{0.242}            & 25.526          & 0.658           & \multicolumn{1}{c|}{0.292}            & 25.655          & 0.804           & \multicolumn{1}{c|}{0.140}            & 23.332          & 0.672           & 0.269
            \\
\multicolumn{1}{l|}{\textbf{EdgeSRGAN-tiny}\distillation[6pt]} & 29.513        & 0.841              & \multicolumn{1}{c|}{0.132}            & 26.950          & 0.727                   & \multicolumn{1}{c|}{0.220}            & 26.174          & 0.673                   & \multicolumn{1}{c|}{0.282}            & 27.106          & 0.845                   & \multicolumn{1}{c|}{0.130}            & 24.117          & 0.704     & 0.249 
\\ \bottomrule
\end{tabular}
}
\caption{Quantitative comparison of different methods for visual-oriented $\times 4$ upsampling. Current SISR state-of-art method SwinIR \cite{liang2021swinir} and bicubic baseline are reported as reference. \scalebox{0.8}{\textbf{↑}}: higher is better, \scalebox{0.8}{\textbf{↓}}: lower is better. $\dag$: trained on DIV2K \cite{agustsson2017ntire} + Flickr2K \cite{timofte2017ntire} + OST \cite{wang2018recovering}.}
\label{tab:SRresGAN}
\end{table}

\begin{table}[t]
\resizebox{\textwidth}{!}{
\begin{tabular}{@{}lcccccccccccccccc@{}}
& & \multicolumn{3}{c}{\textbf{Set5} \cite{bevilacqua2012low}}  & \multicolumn{3}{c}{\textbf{Set14} \cite{zeyde2010single}} & \multicolumn{3}{c}{\textbf{BSD100} \cite{martin2001database}}  & \multicolumn{3}{c}{\textbf{Manga109} \cite{matsui2017sketch}} & \multicolumn{3}{c}{\textbf{Urban100} \cite{huang2015single}} \\ \midrule

\multicolumn{1}{l}{\textbf{Model}}     &  \multicolumn{1}{l|}{}   & \textbf{PSNR ↑} & \textbf{SSIM ↑} & \multicolumn{1}{c|}{\textbf{LPIPS ↓}} & \textbf{PSNR ↑} & \textbf{SSIM ↑} & \multicolumn{1}{c|}{\textbf{LPIPS ↓}} & \textbf{PSNR ↑} & \textbf{SSIM ↑} & \multicolumn{1}{c|}{\textbf{LPIPS ↓}} & \textbf{PSNR ↑} & \textbf{SSIM ↑} & \multicolumn{1}{c|}{\textbf{LPIPS ↓}} & \textbf{PSNR ↑} & \textbf{SSIM ↑} & \textbf{LPIPS ↓} \\ \midrule

\multicolumn{1}{l}{\textbf{Bicubic}}   & \multicolumn{1}{c|}{} & 24.526               & 0.659                & \multicolumn{1}{c|}{0.533} & 23.279               & 0.568                & \multicolumn{1}{c|}{0.628} & 23.727               & 0.546                & \multicolumn{1}{c|}{0.713} & 21.550               & 0.646                & \multicolumn{1}{c|}{0.535} & 20.804               & 0.515                & 0.686                \\

\multicolumn{1}{l}{\textbf{SwinIR} \cite{liang2021swinir}} & \multicolumn{1}{c|}{} & 27.363 & 0.787 & \multicolumn{1}{c|}{0.284} & 25.265 & 0.652 & \multicolumn{1}{c|}{0.428} & 24.984 & 0.606 & \multicolumn{1}{c|}{0.537} & 25.246 & 0.800 & \multicolumn{1}{c|}{0.229} & 23.023 & 0.646 & 0.375 \\ \midrule

\multicolumn{1}{l|}{\textbf{EdgeSRGAN}} & \multicolumn{1}{c|}{\multirow{2}{*}{content}} & 26.462          & 0.755           & \multicolumn{1}{c|}{0.321}            & 24.507          & 0.626           & \multicolumn{1}{c|}{0.460}            & 24.590          & 0.587           & \multicolumn{1}{c|}{0.567}            & 23.840          & 0.753           & \multicolumn{1}{c|}{0.294}            & 22.001          & 0.592           & 0.463            \\

\multicolumn{1}{l|}{\textbf{EdgeSRGAN-tiny}} & \multicolumn{1}{c|}{} & 26.025          & 0.732           & \multicolumn{1}{c|}{0.359}            & 24.286          & 0.615           & \multicolumn{1}{c|}{0.488}            & 24.383          & 0.577           & \multicolumn{1}{c|}{0.591}            & 23.154          & 0.723           & \multicolumn{1}{c|}{0.353}            & 21.680          & 0.570           & 0.520
\\ \midrule

\multicolumn{1}{l|}{\textbf{EdgeSRGAN}} & \multicolumn{1}{c|}{\multirow{2}{*}{visual}} & 25.307          & 0.680           & \multicolumn{1}{c|}{0.228}            & 23.585          & 0.558           & \multicolumn{1}{c|}{0.348}            & 23.547          & 0.514           & \multicolumn{1}{c|}{0.386}            & 22.719          & 0.680           & \multicolumn{1}{c|}{0.257}            & 21.102          & 0.522           & 0.374            \\

\multicolumn{1}{l|}{\textbf{EdgeSRGAN-tiny}} & \multicolumn{1}{c|}{} & 25.523          & 0.693           & \multicolumn{1}{c|}{0.280}            & 23.976          & 0.589           & \multicolumn{1}{c|}{0.399}            & 24.163          & 0.557           & \multicolumn{1}{c|}{0.475}            & 22.874         & 0.695           & \multicolumn{1}{c|}{0.317}            & 21.477          & 0.546           & 0.459
            \\ \bottomrule
\end{tabular}
}
\caption{Quantitative performance of the proposed method for $\times 8$ upsampling. Current SISR state-of-art method SwinIR \cite{liang2021swinir}, and bicubic are reported as references. \scalebox{0.8}{\textbf{↑}}: higher is better, \scalebox{0.8}{\textbf{↓}}: lower is better.}
\label{tab:SRresX8}
\end{table}

\subsection{Super-Resolution Results}
\label{sec:SRresults}
To present quantitative results on image super-resolution, we refer to content-oriented SR for models trained with content-based loss only and visual-oriented SR for models trained with adversarial and perceptual losses. Content-based loss (mean absolute error or mean squared error) aims to maximize PSNR and SSIM, while adversarial and perceptual losses aim to maximize visual quality. We test EdgeSRGAN models on five benchmark datasets (Set5 \cite{bevilacqua2012low}, Set14 \cite{zeyde2010single}, BSD100 \cite{martin2001database}, Manga109 \cite{matsui2017sketch}, and Urban100 \cite{huang2015single}) measuring PSNR, SSIM, and LPIPS. We follow the standard procedure for SISR adopted in \cite{liang2021swinir}, where the metrics are computed on the luminance channel Y of the YCbCr converted images. Also, $S$ pixels are cropped from each image border, where $S$ is the model scale factor.

\begin{figure*}[!htbp]
\begin{subfigure}[!htbp]{1.00\linewidth}
    \captionsetup{font=small}
    \centering
	\scriptsize
	
	\newcommand{\h}{0.105}
	\newcommand{\wa}{0.12}
	\newcommand{\wb}{0.16}
	\renewcommand{\g}{-0.7mm}
	\renewcommand{\tabcolsep}{1.8pt}
	\renewcommand{\arraystretch}{1}
	\resizebox{\linewidth}{!} {
		\begin{tabular}{cc}
			
			\newcommand{\name}{img003_}
			\renewcommand{\h}{0.078}
			\newcommand{\w}{0.176}
			\begin{tabular}{cc}
				\begin{adjustbox}{valign=t}
					\begin{tabular}{c}
						\includegraphics[height=0.185\textwidth]{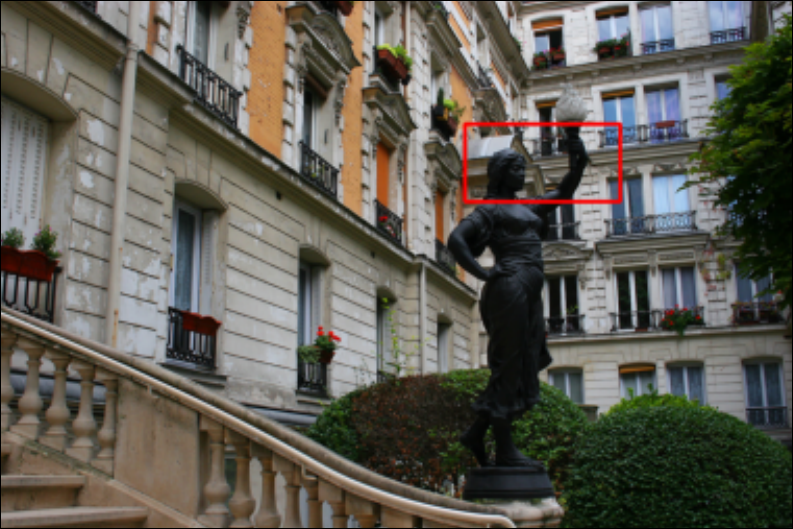}
						\\
						Urban100 ($\times 4$): \tt{img\_003}
					\end{tabular}
				\end{adjustbox}
				\begin{adjustbox}{valign=t}
					\begin{tabular}{cccccc}
						\includegraphics[height=\h \textwidth, width=\w \textwidth]{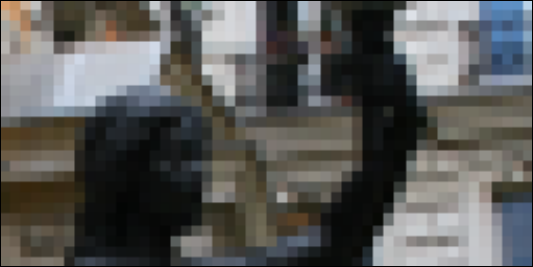} \hspace{\g} &
						\includegraphics[height=\h \textwidth, width=\w \textwidth]{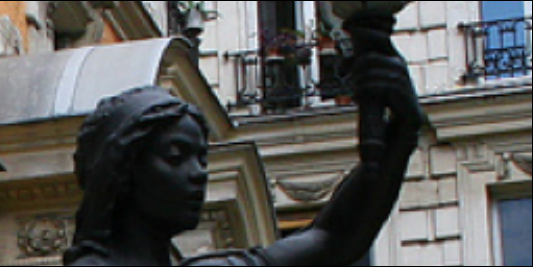} \hspace{\g} &
						\includegraphics[height=\h \textwidth, width=\w \textwidth]{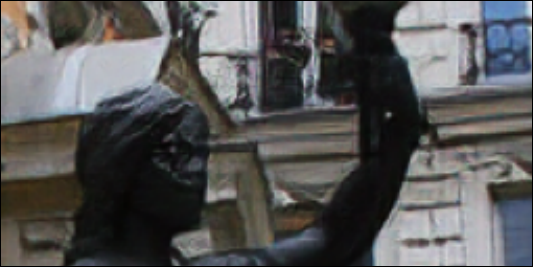} \hspace{\g} &
						\includegraphics[height=\h \textwidth, width=\w \textwidth]{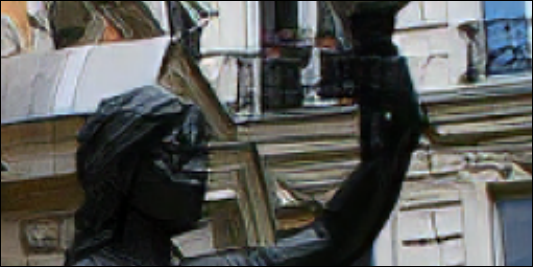}
						\\
						LR \hspace{\g} &
						HR \hspace{\g} &
						SRGAN~\cite{ledig2017photo} \hspace{\g} &
						ESRGAN~\cite{wang2018esrgan} \hspace{\g}
						\\
						\vspace{-1.5mm}
						\\
						
						\includegraphics[height=\h \textwidth, width=\w \textwidth]{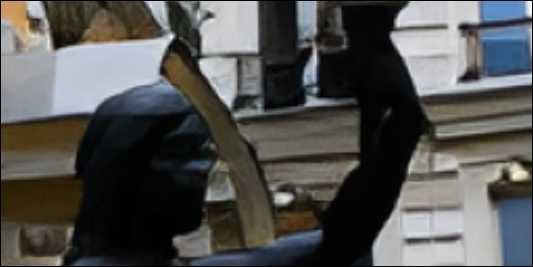} \hspace{\g} &
						\includegraphics[height=\h \textwidth, width=\w \textwidth]{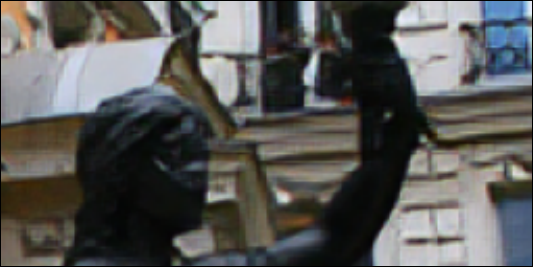} \hspace{\g} &
						\includegraphics[height=\h \textwidth, width=\w \textwidth]{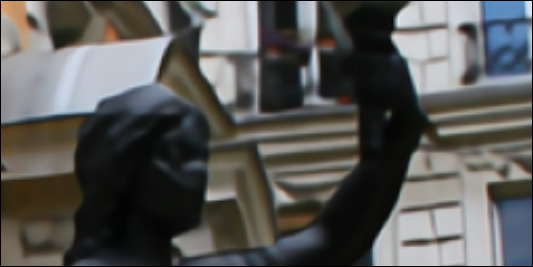} \hspace{\g} &		
						\includegraphics[height=\h \textwidth, width=\w \textwidth]{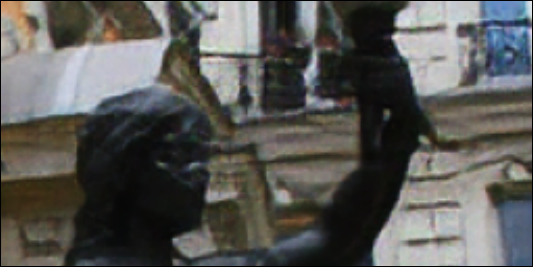}
						\\ 
						
						RealESRGAN~\cite{chen2022real} \hspace{\g} &
						AGD~\cite{fu2020autogan} \hspace{\g} &
						SwinIR~\cite{liang2021swinir} \hspace{\g} &
					    \textbf{EdgeSRGAN (ours)} \hspace{\g}
						\\
					\end{tabular}
				\end{adjustbox}
			\end{tabular}
			
		\end{tabular}
	}\vspace{-2mm}
	\label{fig:visual_urban100}
\end{subfigure}
\vskip 12pt
\begin{subfigure}[!htbp]{1.00\linewidth}
    \captionsetup{font=small}
    \centering
	\scriptsize
	
	\newcommand{\h}{0.105}
	\newcommand{\wa}{0.12}
	\newcommand{\wb}{0.16}
	\renewcommand{\g}{-0.7mm}
	\renewcommand{\tabcolsep}{1.8pt}
	\renewcommand{\arraystretch}{1}
	\resizebox{\linewidth}{!} {
		\begin{tabular}{cc}
			
			\newcommand{\name}{ParaisoRoad_}
			\renewcommand{\h}{0.078}
			\newcommand{\w}{0.176}
			\begin{tabular}{cc}
				\begin{adjustbox}{valign=t}
					\begin{tabular}{c}
						\includegraphics[height=0.185\textwidth]{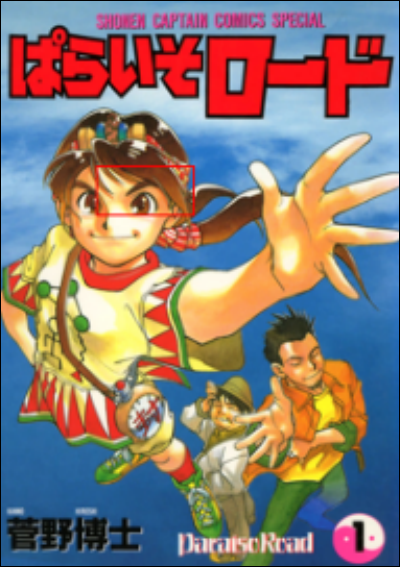}
						\\
						Manga109 ($\times 4$): \tt{ParaisoRoad}
					\end{tabular}
				\end{adjustbox}
				\begin{adjustbox}{valign=t}
					\begin{tabular}{cccccc}
						\includegraphics[height=\h \textwidth, width=\w \textwidth]{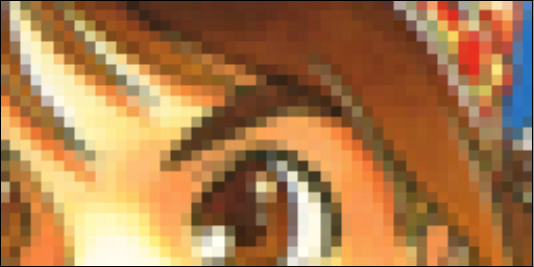} \hspace{\g} &
						\includegraphics[height=\h \textwidth, width=\w \textwidth]{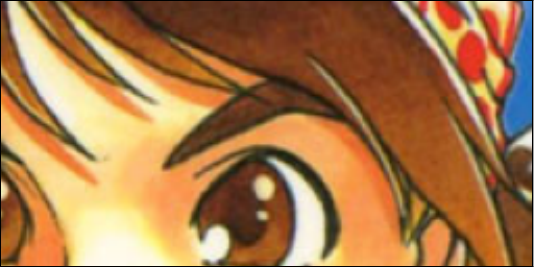} \hspace{\g} &
						\includegraphics[height=\h \textwidth, width=\w \textwidth]{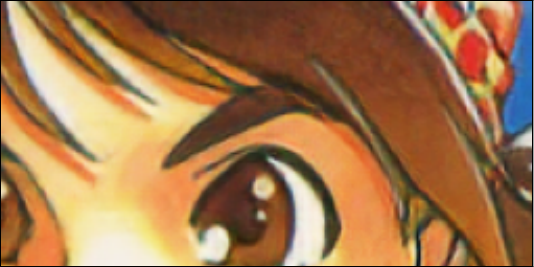} \hspace{\g} &
						\includegraphics[height=\h \textwidth, width=\w \textwidth]{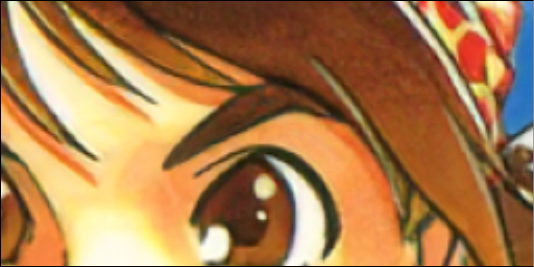}
						\\
						LR \hspace{\g} &
						HR \hspace{\g} &
						SRGAN~\cite{ledig2017photo} \hspace{\g} &
						ESRGAN~\cite{wang2018esrgan} \hspace{\g}
						\\
						\vspace{-1.5mm}
						\\
						
						\includegraphics[height=\h \textwidth, width=\w \textwidth]{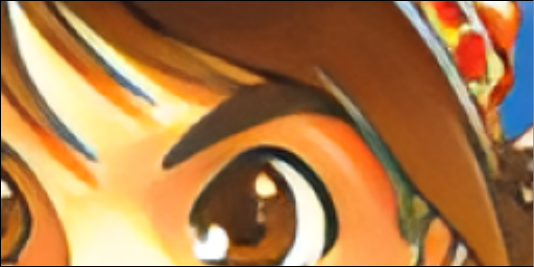} \hspace{\g} &
						\includegraphics[height=\h \textwidth, width=\w \textwidth]{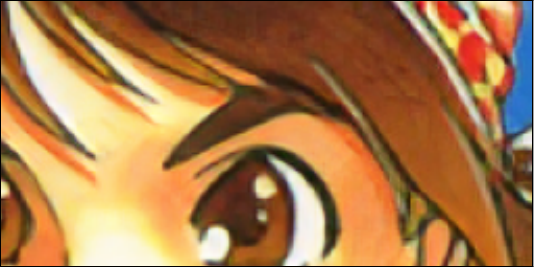} \hspace{\g} &
						\includegraphics[height=\h \textwidth, width=\w \textwidth]{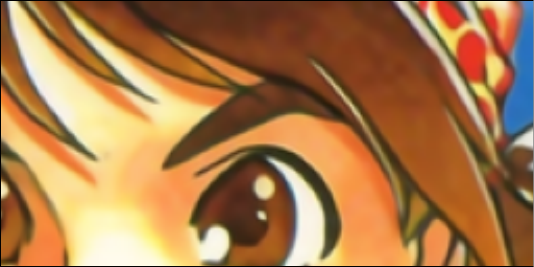} \hspace{\g} &		
						\includegraphics[height=\h \textwidth, width=\w \textwidth]{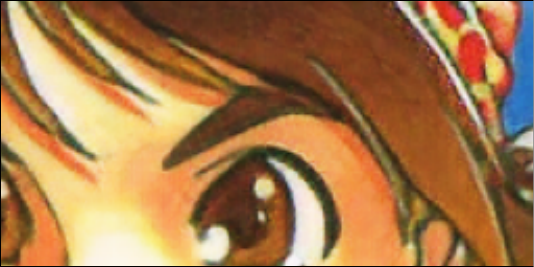}
						\\ 
						
						RealESRGAN~\cite{chen2022real} \hspace{\g} &
						AGD~\cite{fu2020autogan} \hspace{\g} &
						SwinIR~\cite{liang2021swinir} \hspace{\g} &
					    \textbf{EdgeSRGAN (ours)} \hspace{\g}
						\\
					\end{tabular}
				\end{adjustbox}
			\end{tabular}
			
		\end{tabular}
	}\vspace{-2mm}
	\label{fig:visual_manga100}
\end{subfigure}
\vskip 12pt
\begin{subfigure}[!htbp]{1.00\linewidth}
    \captionsetup{font=small}
    \centering
	\scriptsize
	
	\newcommand{\h}{0.105}
	\newcommand{\wa}{0.12}
	\newcommand{\wb}{0.16}
	\renewcommand{\g}{-0.7mm}
	\renewcommand{\tabcolsep}{1.8pt}
	\renewcommand{\arraystretch}{1}
	\resizebox{\linewidth}{!} {
		\begin{tabular}{cc}
			
			\newcommand{\name}{108070_}
			\renewcommand{\h}{0.078}
			\newcommand{\w}{0.176}
			\begin{tabular}{cc}
				\begin{adjustbox}{valign=t}
					\begin{tabular}{c}
						\includegraphics[height=0.185\textwidth]{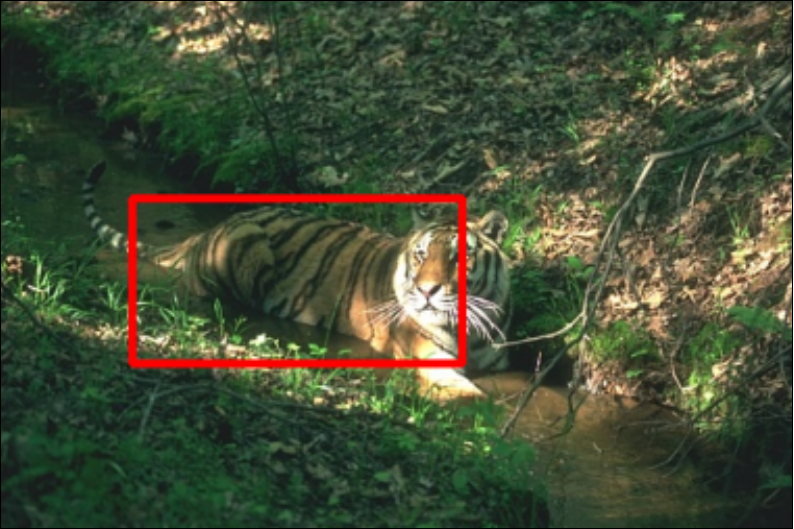}
						\\
						BSD100 ($\times 4$): \tt{108070}
					\end{tabular}
				\end{adjustbox}
				\begin{adjustbox}{valign=t}
					\begin{tabular}{cccccc}
						\includegraphics[height=\h \textwidth, width=\w \textwidth]{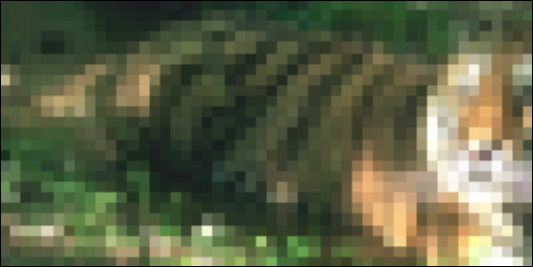} \hspace{\g} &
						\includegraphics[height=\h \textwidth, width=\w \textwidth]{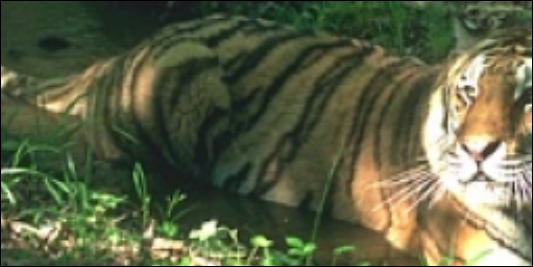} \hspace{\g} &
						\includegraphics[height=\h \textwidth, width=\w \textwidth]{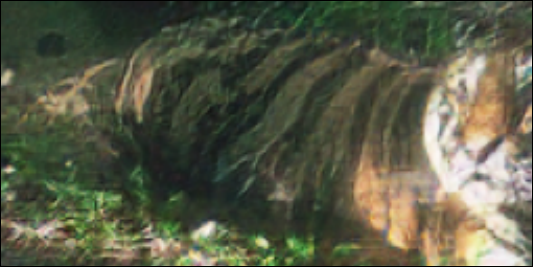} \hspace{\g} &
						\includegraphics[height=\h \textwidth, width=\w \textwidth]{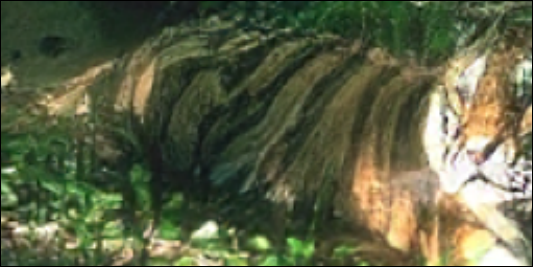}
						\\
						LR \hspace{\g} &
						HR \hspace{\g} &
						SRGAN~\cite{ledig2017photo} \hspace{\g} &
						ESRGAN~\cite{wang2018esrgan} \hspace{\g}
						\\
						\vspace{-1.5mm}
						\\
						
						\includegraphics[height=\h \textwidth, width=\w \textwidth]{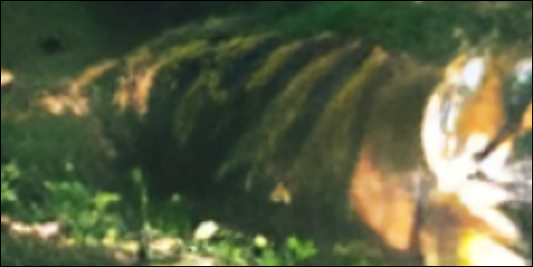} \hspace{\g} &
						\includegraphics[height=\h \textwidth, width=\w \textwidth]{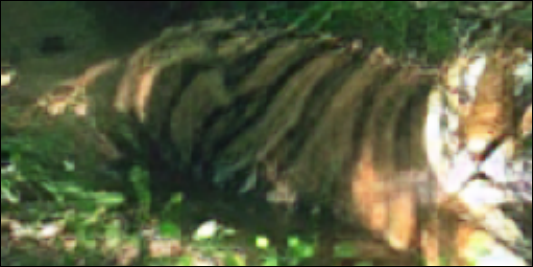} \hspace{\g} &
						\includegraphics[height=\h \textwidth, width=\w \textwidth]{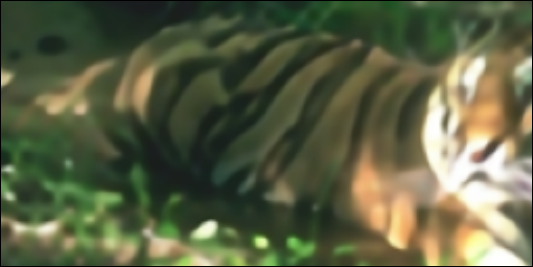} \hspace{\g} &		
						\includegraphics[height=\h \textwidth, width=\w \textwidth]{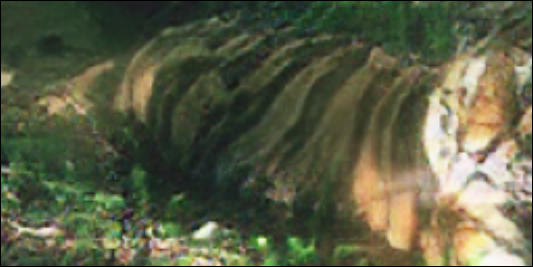}
						\\ 
						
						RealESRGAN~\cite{chen2022real} \hspace{\g} &
						AGD~\cite{fu2020autogan} \hspace{\g} &
						SwinIR~\cite{liang2021swinir} \hspace{\g} &
					    \textbf{EdgeSRGAN (ours)} \hspace{\g}
						\\
					\end{tabular}
				\end{adjustbox}
			\end{tabular}
			
		\end{tabular}
	}\vspace{-2mm}
	\label{fig:visual_bsd100}
\end{subfigure}
\vskip 12pt
\begin{subfigure}[!htbp]{1.00\linewidth}
    \captionsetup{font=small}
    \centering
	\scriptsize
	
	\newcommand{\h}{0.105}
	\newcommand{\wa}{0.12}
	\newcommand{\wb}{0.16}
	\renewcommand{\g}{-0.7mm}
	\renewcommand{\tabcolsep}{1.8pt}
	\renewcommand{\arraystretch}{1}
	\resizebox{\linewidth}{!} {
		\begin{tabular}{cc}
			
			\newcommand{\name}{butterfly_}
			\renewcommand{\h}{0.078}
			\newcommand{\w}{0.176}
			\begin{tabular}{cc}
				\begin{adjustbox}{valign=t}
					\begin{tabular}{c}
						\includegraphics[height=0.185\textwidth]{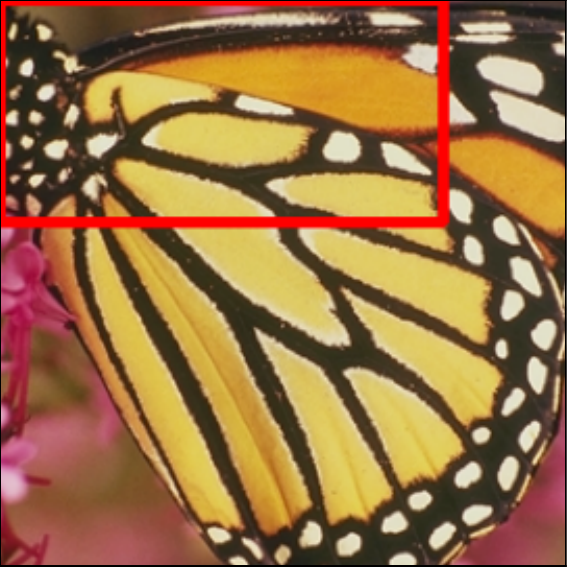}
						\\
						Set5 ($\times 4$): \tt{butterfly}
					\end{tabular}
				\end{adjustbox}
				\begin{adjustbox}{valign=t}
					\begin{tabular}{cccccc}
						\includegraphics[height=\h \textwidth, width=\w \textwidth]{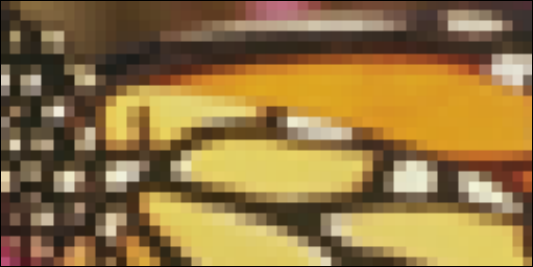} \hspace{\g} &
						\includegraphics[height=\h \textwidth, width=\w \textwidth]{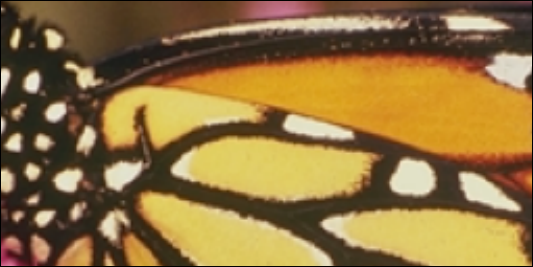} \hspace{\g} &
						\includegraphics[height=\h \textwidth, width=\w \textwidth]{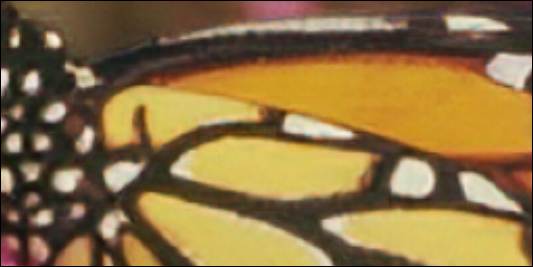} \hspace{\g} &
						\includegraphics[height=\h \textwidth, width=\w \textwidth]{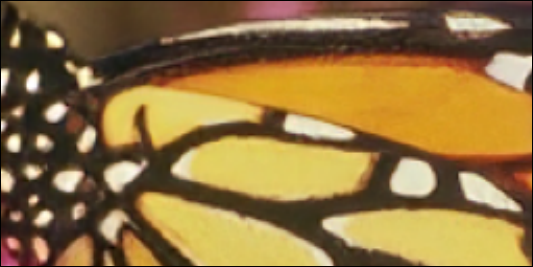}
						\\
						LR \hspace{\g} &
						HR \hspace{\g} &
						SRGAN~\cite{ledig2017photo} \hspace{\g} &
						ESRGAN~\cite{wang2018esrgan} \hspace{\g}
						\\
						\vspace{-1.5mm}
						\\
						
						\includegraphics[height=\h \textwidth, width=\w \textwidth]{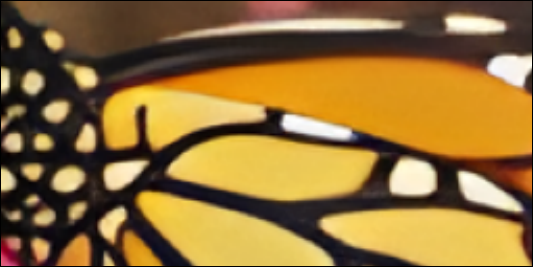} \hspace{\g} &
						\includegraphics[height=\h \textwidth, width=\w \textwidth]{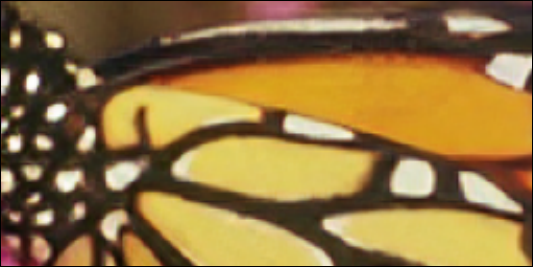} \hspace{\g} &
						\includegraphics[height=\h \textwidth, width=\w \textwidth]{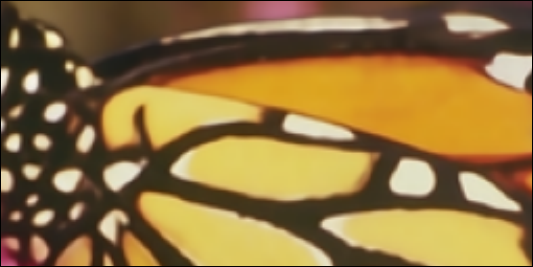} \hspace{\g} &		
						\includegraphics[height=\h \textwidth, width=\w \textwidth]{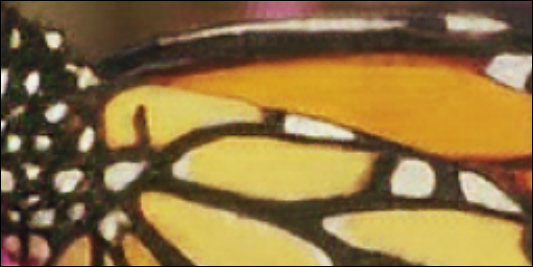}
						\\ 
						
						RealESRGAN~\cite{chen2022real} \hspace{\g} &
						AGD~\cite{fu2020autogan} \hspace{\g} &
						SwinIR~\cite{liang2021swinir} \hspace{\g} &
					    \textbf{EdgeSRGAN (ours)} \hspace{\g}
						\\
					\end{tabular}
				\end{adjustbox}
			\end{tabular}
			
		\end{tabular}
	}\vspace{-2mm}
	\label{fig:visual_set5}
\end{subfigure}
\vskip 12pt
\begin{subfigure}[!htbp]{1.00\linewidth}
    \captionsetup{font=small}
    \centering
	\scriptsize
	
	\newcommand{\h}{0.105}
	\newcommand{\wa}{0.12}
	\newcommand{\wb}{0.16}
	\renewcommand{\g}{-0.7mm}
	\renewcommand{\tabcolsep}{1.8pt}
	\renewcommand{\arraystretch}{1}
	\resizebox{\linewidth}{!} {
		\begin{tabular}{cc}
			
			\newcommand{\name}{baboon_}
			\renewcommand{\h}{0.078}
			\newcommand{\w}{0.176}
			\begin{tabular}{cc}
				\begin{adjustbox}{valign=t}
					\begin{tabular}{c}
						\includegraphics[height=0.185\textwidth]{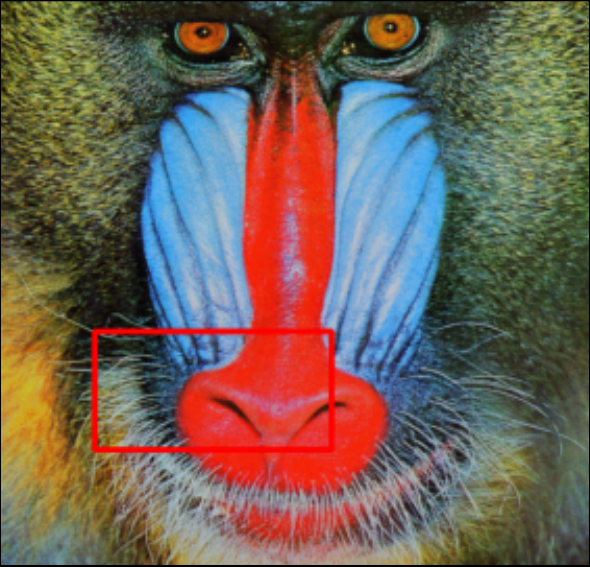}
						\\
						Set14 ($\times 4$): \tt{baboon}
					\end{tabular}
				\end{adjustbox}
				\begin{adjustbox}{valign=t}
					\begin{tabular}{cccccc}
						\includegraphics[height=\h \textwidth, width=\w \textwidth]{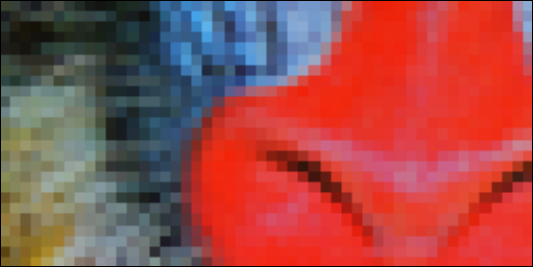} \hspace{\g} &
						\includegraphics[height=\h \textwidth, width=\w \textwidth]{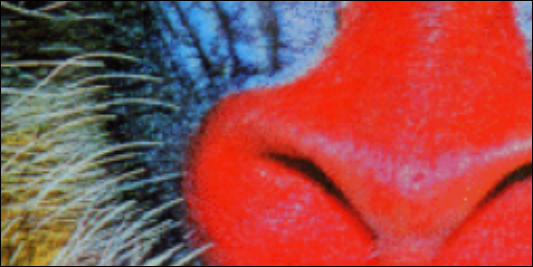} \hspace{\g} &
						\includegraphics[height=\h \textwidth, width=\w \textwidth]{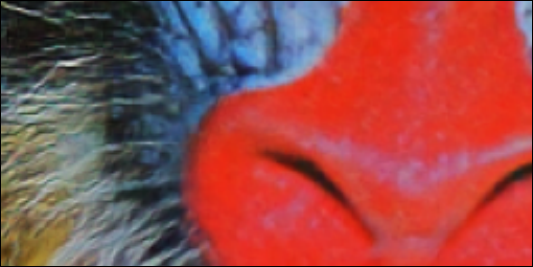} \hspace{\g} &
						\includegraphics[height=\h \textwidth, width=\w \textwidth]{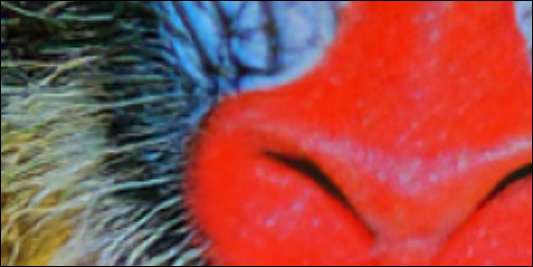}
						\\
						LR \hspace{\g} &
						HR \hspace{\g} &
						SRGAN~\cite{ledig2017photo} \hspace{\g} &
						ESRGAN~\cite{wang2018esrgan} \hspace{\g}
						\\
						\vspace{-1.5mm}
						\\
						
						\includegraphics[height=\h \textwidth, width=\w \textwidth]{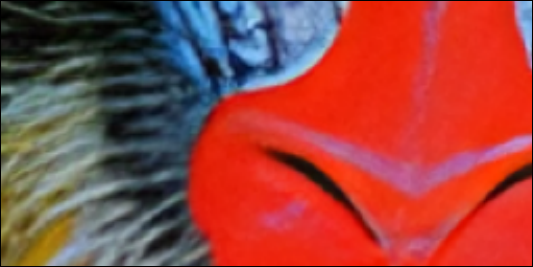} \hspace{\g} &
						\includegraphics[height=\h \textwidth, width=\w \textwidth]{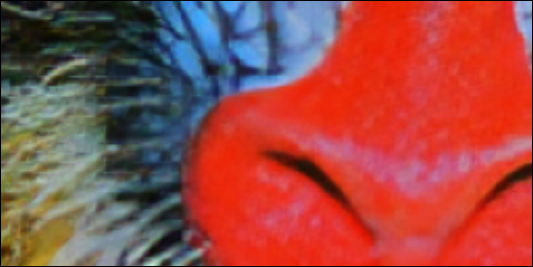} \hspace{\g} &
						\includegraphics[height=\h \textwidth, width=\w \textwidth]{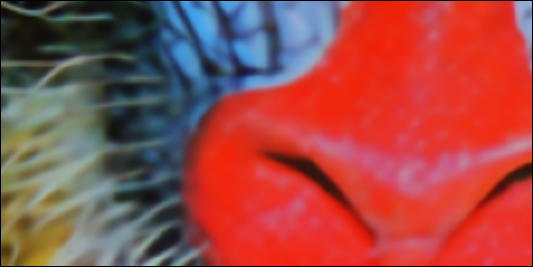} \hspace{\g} &		
						\includegraphics[height=\h \textwidth, width=\w \textwidth]{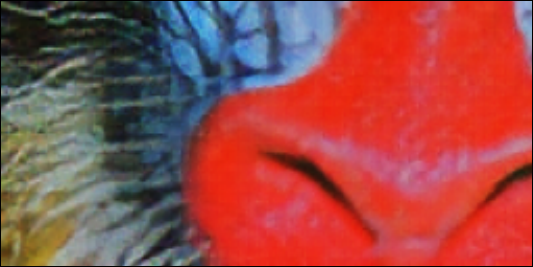}
						\\ 
						
						RealESRGAN~\cite{chen2022real} \hspace{\g} &
						AGD~\cite{fu2020autogan} \hspace{\g} &
						SwinIR~\cite{liang2021swinir} \hspace{\g} &
					    \textbf{EdgeSRGAN (ours)} \hspace{\g}
						\\
					\end{tabular}
				\end{adjustbox}
			\end{tabular}
			
		\end{tabular}
	}\vspace{-2mm}
	\label{fig:visual_mset14}
\end{subfigure}
\vskip 8pt
\caption{Visual comparison of bicubic image SR ($\times 4$) methods on random samples from the considered datasets. EdgeSRGAN achieves results that are comparable to state-of-the-art solutions with $\sim10\%$ of the weights.}
\label{fig:visual_comparison}
\end{figure*}

Tab. \ref{tab:SRresPSNR} and Tab. \ref{tab:SRresGAN} show the comparison with other methods for content-oriented and visual-oriented $\times 4$ SR, respectively. We report results of other GAN-based methodologies \cite{ledig2017photo,wang2018esrgan,wang2021real-esrgan,fu2020autogan} as well as the current content-oriented SOTA SwinIR \cite{liang2021swinir} and bicubic baseline, as reference. Unlike what is usually found in literature, we refer to the OpenCV\footnote{\url{https://docs.opencv.org/2.4/modules/imgproc/doc/geometric\_transformations.html\#resize}} bicubic resize implementation instead of the one present in MATLAB. For visual-oriented SR, we also report the results of the distilled tiny model EdgeSRGAN-tiny\distillation[6pt]. The proposed method reaches competitive results in all the metrics, even with some degradation for tiny models due to the considerable weight reduction. The distillation method helps EdgeSRGAN-tiny training by transferring knowledge from the standard model and decreasing the degradation due to the reduced number of parameters. Note that ESRGAN and RealESRGAN are trained on Flickr2K \cite{timofte2017ntire}, and OST \cite{wang2018recovering} datasets in addition to DIV2K.
Tab. \ref{tab:SRresX8} reports results of the $\times 8$ models, together with SwinIR and bicubic. Also, in this case, the proposed models reach competitive results, and knowledge distillation helps to reduce performance degradation in the tiny model.
As a final qualitative evaluation, Fig. \ref{fig:visual_comparison} compares the super-resolved images obtained by EdgeSRGAN with the considered state-of-the-art solutions. Our model shows comparable results, highlighting more texture and details than networks trained with pixel loss ($L_\text{MSE}$) while remaining true to the ground truth image. 

\begin{figure*}[!htbp]
    \captionsetup{font=small}
    \centering
	\scriptsize
	
	\newcommand{\h}{0.105}
	\newcommand{\wa}{0.12}
	\newcommand{\wb}{0.16}
	\renewcommand{\g}{-0.7mm}
	\renewcommand{\tabcolsep}{1.8pt}
	\renewcommand{\arraystretch}{1}
	\resizebox{\linewidth}{!} {
		\begin{tabular}{cc}
			
			\newcommand{\name}{baby_}
			\renewcommand{\h}{0.078}
			\newcommand{\w}{0.176}
			\begin{tabular}{cc}
				\begin{adjustbox}{valign=t}
					\begin{tabular}{c}
						\includegraphics[height=0.185\textwidth]{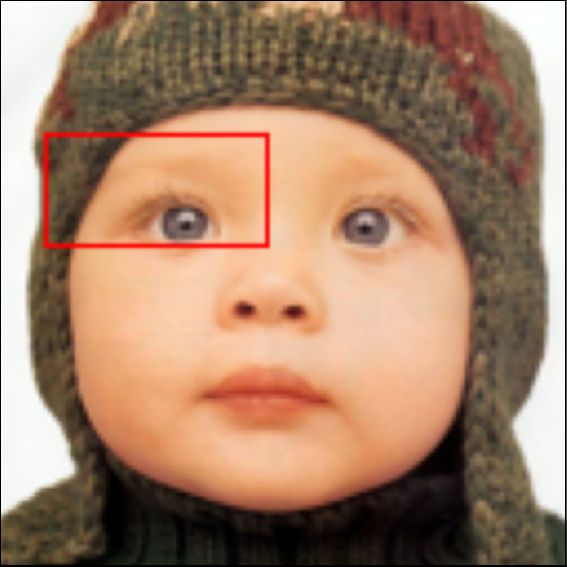}
						\\
						Set5 ($\times 4$): \tt{baby}
					\end{tabular}
				\end{adjustbox}
				\begin{adjustbox}{valign=t}
					\begin{tabular}{cccccc}
						\includegraphics[height=\h \textwidth, width=\w \textwidth]{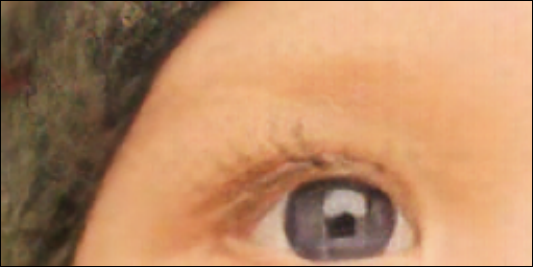} \hspace{\g} &
						\includegraphics[height=\h \textwidth, width=\w \textwidth]{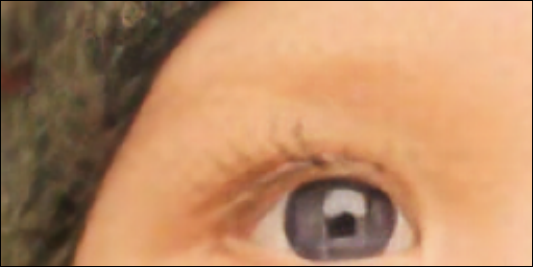} \hspace{\g} &
						\includegraphics[height=\h \textwidth, width=\w \textwidth]{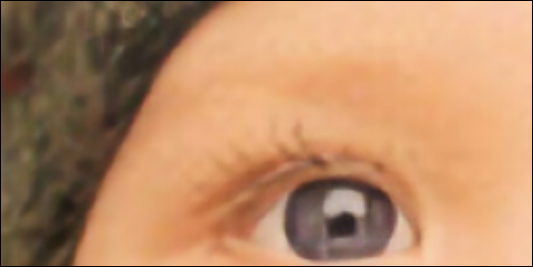}
						\\
						$\alpha = 0$ \hspace{\g} &
						$\alpha = 0.2$ \hspace{\g} &
						$\alpha = 0.4$ \hspace{\g}
						\\
						\vspace{-1.5mm}
						\\
						
						\includegraphics[height=\h \textwidth, width=\w \textwidth]{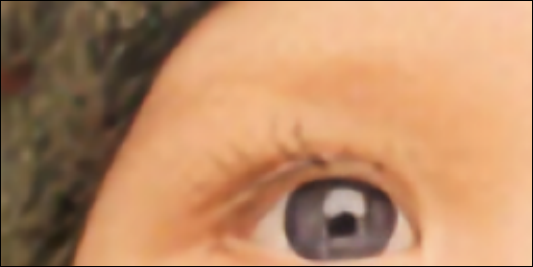} \hspace{\g} &
						\includegraphics[height=\h \textwidth, width=\w \textwidth]{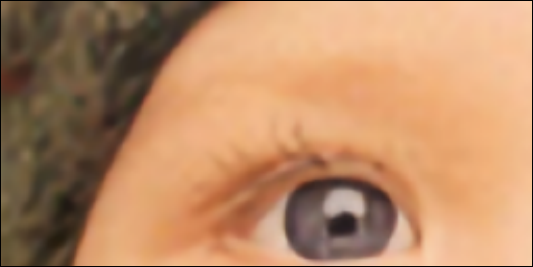} \hspace{\g} &
						\includegraphics[height=\h \textwidth, width=\w \textwidth]{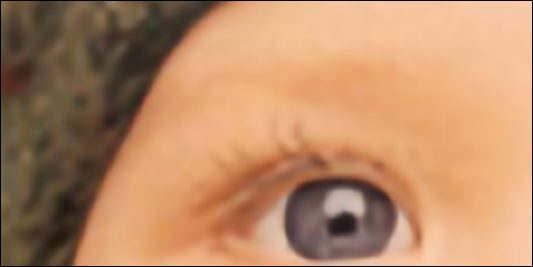}
						\\ 
						
						$\alpha = 0.6$ \hspace{\g} &
						$\alpha = 0.8$ \hspace{\g} &
					    $\alpha = 1$ \hspace{\g}
						\\
					\end{tabular}
				\end{adjustbox}
			\end{tabular}
			
		\end{tabular}
	}\vspace{-2mm}
	\caption{Visual comparison of interpolated EdgeSRGAN for different values of $\alpha$. Values closer to 1 generate outputs focused on content fidelity, while small values go towards visually pleasing results.}
	\label{fig:visual_interpolation}
\end{figure*}

\subsubsection{Model Interpolation}
\label{sec:NIresults}
\begin{figure}[t]
    \centering
    \begin{subfigure}{\linewidth}
        \centering
        \includegraphics[width=\textwidth]{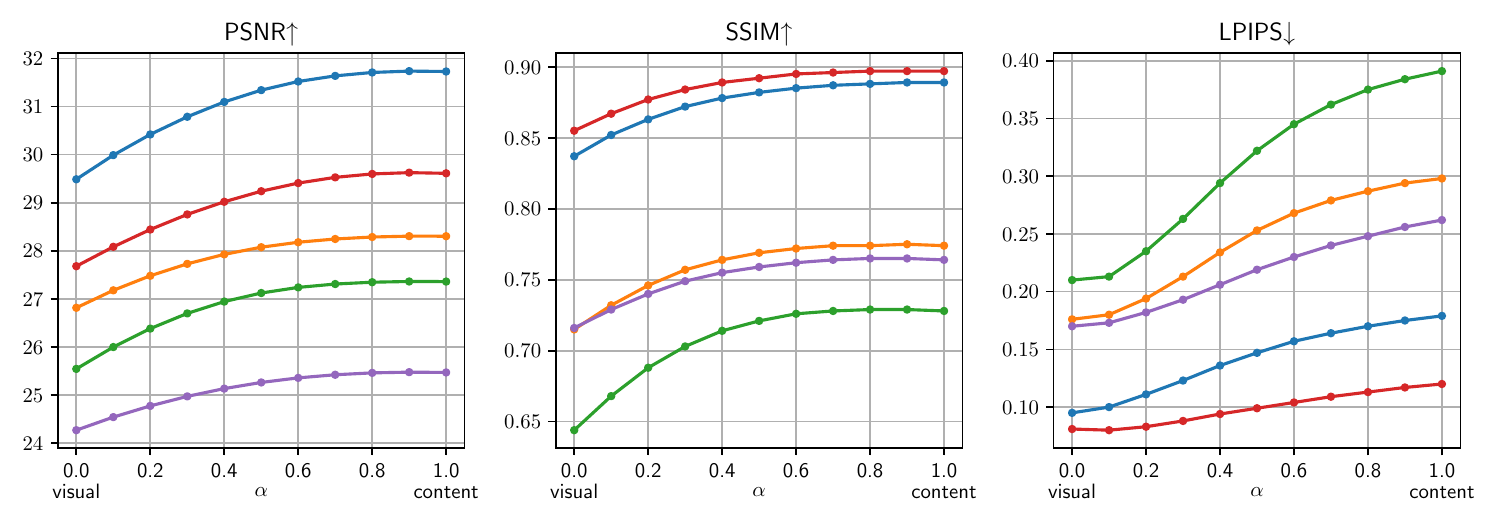}
    \end{subfigure}
    \begin{subfigure}{\linewidth}
        \centering
        \includegraphics[width=0.45\columnwidth, trim={10px 10px 10px 10px}, clip]{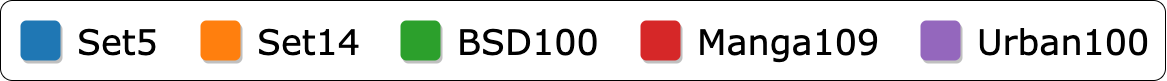}
    \end{subfigure}
	\caption{EdgeSRGAN network interpolation results on the benchmark datasets for $\times 4$ upsampling. Changing the network interpolation weight $\alpha$, it is possible to select the desired trade-off between content-oriented and visual-oriented SR. \protect\newline
	\scalebox{0.8}{\textbf{↑}}: higher is better, \scalebox{0.8}{\textbf{↓}}: lower is better.}\label{fig:interpolation}
\end{figure}

 We report the results of network interpolation on the benchmark datasets in Fig. \ref{fig:interpolation}. We consider $\alpha$ values between 0 and 1 with a step of 0.1, with 0 implying a full visual-oriented model and 1 a full content-oriented one. All results refer to the standard EdgeSRGAN model for $\times 4$ upsampling. This procedure effectively shows how it is possible to choose the desired trade-off between content-oriented and visual-oriented SR simply by changing the interpolation weight $\alpha$. An increase in the weight value causes an improvement of the content-related metrics PSNR and SSIM and a worsening of the perceptual index LPIPS. This behavior holds for all the test datasets, validating the proposed approach. This procedure can be easily carried out in a real-time application and only requires computing the interpolated weights once. Thus, it does not affect any way the inference speed. For an additional visual evaluation, Fig. \ref{fig:visual_interpolation} reports the outputs obtained for increasing values of $\alpha$ on a random dataset sample.

\subsubsection{Model Quantization} 
\label{sec:NOresults}
To target Edge TPU devices and reach over-real-time inference results, we follow the quantization scheme of Eq. \ref{eq:quantization} for both weights and activations to obtain a full-integer model. Since quantized models must have a fixed input shape, we generate a full-integer network for each input shape of the testing samples. We use the 100 images from the DIV2K validation set as a representative dataset to calibrate the quantization algorithm. We refer to the int8-quantized standard model as EdgeSRGANi8. As for the tiny model, we optimize the distilled network EdgeSRGANi8-tiny\distillation[6pt]. Results for the visual-oriented optimized models are shown in Tab. \ref{tab:SRresCoral}. Due to the full-integer models' reduced activation and weight, we experience a great increase in inference speed up to over-real-time at the cost of degradation in SR performance. All the proposed quantized models still outperform the bicubic baseline on the perceptual index LPIPS and therefore represent a good option for applications in which really fast inference is needed. A comparison of different models for visual-oriented $\times 4$ upsampling is shown in Fig. \ref{fig:fps_lpips}. We consider LPIPS performance on the Set5 dataset compared to framerate.

\begin{table}[ht]
\resizebox{\textwidth}{!}{
\begin{tabular}{@{}lcccccccccccccccc@{}}
& & \multicolumn{3}{c}{\textbf{Set5} \cite{bevilacqua2012low}}  & \multicolumn{3}{c}{\textbf{Set14} \cite{zeyde2010single}} & \multicolumn{3}{c}{\textbf{BSD100} \cite{martin2001database}}  & \multicolumn{3}{c}{\textbf{Manga109} \cite{matsui2017sketch}} & \multicolumn{3}{c}{\textbf{Urban100} \cite{huang2015single}} \\ \midrule

\multicolumn{1}{l}{\textbf{Model}}     &  \multicolumn{1}{l|}{\textbf{Scale}}   & \textbf{PSNR ↑} & \textbf{SSIM ↑} & \multicolumn{1}{c|}{\textbf{LPIPS ↓}} & \textbf{PSNR ↑} & \textbf{SSIM ↑} & \multicolumn{1}{c|}{\textbf{LPIPS ↓}} & \textbf{PSNR ↑} & \textbf{SSIM ↑} & \multicolumn{1}{c|}{\textbf{LPIPS ↓}} & \textbf{PSNR ↑} & \textbf{SSIM ↑} & \multicolumn{1}{c|}{\textbf{LPIPS ↓}} & \textbf{PSNR ↑} & \textbf{SSIM ↑} & \textbf{LPIPS ↓} \\ \midrule

\multicolumn{1}{l|}{\textbf{EdgeSRGANi8}} & \multicolumn{1}{c|}{\multirow{2}{*}{$\times4$}} & 27.186          & 0.721           & \multicolumn{1}{c|}{0.209}            & 24.714          & 0.475           & \multicolumn{1}{c|}{0.342}            & 23.675          & 0.484           & \multicolumn{1}{c|}{0.438}            & 25.601          & 0.712           & \multicolumn{1}{c|}{0.221}            & 22.802          & 0.580           & 0.341            \\

\multicolumn{1}{l|}{\textbf{EdgeSRGANi8-tiny\distillation}} & \multicolumn{1}{c|}{} & 27.330          & 0.710           & \multicolumn{1}{c|}{0.257}            & 24.807          & 0.562           & \multicolumn{1}{c|}{0.390}            & 23.837          & 0.485           & \multicolumn{1}{c|}{0.481}            & 25.299          & 0.696           & \multicolumn{1}{c|}{0.286}            & 22.580          & 0.538           & 0.454
            \\ \midrule

\multicolumn{1}{l|}{\textbf{EdgeSRGANi8}} & \multicolumn{1}{c|}{\multirow{2}{*}{$\times 8$}} & 24.433          & 0.602          & \multicolumn{1}{c|}{0.312}            & 22.846          & 0.477           & \multicolumn{1}{c|}{0.440}            & 22.609          & 0.422           & \multicolumn{1}{c|}{0.492}            & 22.227    & 0.603  & \multicolumn{1}{c|}{0.342}    & 20.525  & 0.433 &  0.499        \\

\multicolumn{1}{l|}{\textbf{EdgeSRGANi8-tiny}} & \multicolumn{1}{c|}{} & 24.956  & 0.642  & \multicolumn{1}{c|}{0.333} & 23.487  & 0.532   & \multicolumn{1}{c|}{0.461}    & 23.591  & 0.494   & \multicolumn{1}{c|}{0.544}    & 22.445  &  0.632  & \multicolumn{1}{c|}{0.386}   & 21.125 & 0.489  & 0.548 
\\ \bottomrule
\end{tabular}
}
\caption{Quantitative performance of the full-integer quantized  models for $\times 4$ and $\times 8$ visual-based SR. \scalebox{0.8}{\textbf{↑}}: higher is better, \scalebox{0.8}{\textbf{↓}}: lower is better.}
\label{tab:SRresCoral}
\end{table}

\subsection{Ablation Study}
\label{sec:ablation}
To further verify the effectiveness of our model for real-time super-resolution, we conduct an ablation study to analyze the effect of our architectural design choices. In particular, we benchmark EdgeSRGAN at four progressive steps, reporting fidelity, perceptual performance, and inference speed. The steps we consider are the following:
\begin{enumerate}
    \item Reducing the number of residual blocks $N$;
    \item Replacing the Pixel Shuffle upsampling stage with Transpose Convolutions;
    \item Removing Batch Normalization;
    \item Replacing PReLU activations with ReLU.
\end{enumerate}

The last step corresponds to the final version of EdgeSRGAN. For each step of the model, we use the same training procedure described in \ref{sec:training} and measure the inference speed on the CPU at (80x60) and (160x120) input resolutions. All the results are reported in Tab. \ref{tab:ablation}. The experimentation confirms that each compression step gains substantial inference speed by trading minimal perceptual quality. Overall, we observe -3.7\% LPIPS perceptual quality and +280\% inference speed.

\begin{table}[ht]
\centering
\resizebox{\columnwidth}{!}{%
\begin{tabular}{@{}cc|ccc|ccc|ccc|ccc|ccc|cc@{}}
\toprule
\multirow{2}{*}{\textbf{Model}} & \multirow{2}{*}{\textbf{Params}} & \multicolumn{3}{c|}{\textbf{Set5}} & \multicolumn{3}{c|}{\textbf{Set14}} & \multicolumn{3}{c|}{\textbf{BSD100}} & \multicolumn{3}{c|}{\textbf{Manga100}} & \multicolumn{3}{c|}{\textbf{Urban100}} & \multicolumn{2}{c}{\textbf{Inference Speed (fps)}} \\ \cmidrule(l){3-19} 
 &  & \textbf{PSNR ↑} & \textbf{SSIM ↑} & \textbf{LPIPS ↓} & \textbf{PSNR ↑} & \textbf{SSIM ↑} & \textbf{LPIPS ↓} & \textbf{PSNR ↑} & \textbf{SSIM ↑} & \textbf{LPIPS ↓} & \textbf{PSNR ↑} & \textbf{SSIM ↑} & \textbf{LPIPS ↓} & \textbf{PSNR ↑} & \textbf{SSIM ↑} & \textbf{LPIPS ↓} & \textbf{80x60} & \textbf{160x120} \\ \midrule
\textbf{SRGAN} & 1.5M & 29,18 & 0,842 & 0,094 & 26,17 & 0,701 & 0,172 & 25,45 & 0,648 & 0,206 & 27,35 & 0,860 & 0,076 & 24,39 & 0,728 & 0,158 & 2.00 ± 0.03 & 0.48 ± 0.01 \\
\textbf{$N=8$} & 956k & 29,38 & 0,839 & 0,088 & 26,55 & 0,703 & 0,170 & 25,08 & 0,628 & 0,207 & 27,49 & 0,852 & 0,085 & 24,21 & 0,718 & 0,168 & 2.47 ± 0.01 & 0.62 ± 0.01 \\
\textbf{TransposeConv} & 663k & 28,98 & 0,829 & 0,113 & 26,46 & 0,706 & 0,204 & 25,25 & 0,641 & 0,243 & 26,72 & 0,833 & 0,116 & 23,66 & 0,689 & 0,214 & 9.16 ± 0.31 & 2.52 ± 0.03 \\
\textbf{No BatchNorm} & 661k & 29,40 & 0,838 & 0,105 & 26,65 & 0,709 & 0,194 & 25,09 & 0,630 & 0,236 & 27,54 & 0,851 & 0,091 & 24,01 & 0,707 & 0,191 & 9.91 ± 0.16 & 2.56 ± 0.06 \\
\textbf{ReLU} & 661k & 29,49 & 0,837 & 0,095 & 26,81 & 0,715 & 0,176 & 25,54 & 0,644 & 0,210 & 27,68 & 0,855 & 0,081 & 24,27 & 0,716 & 0,170 & 10.26 ± 0.11 & 2.66 ± 0.02 \\ \bottomrule
\end{tabular}%
}
\caption{Results of the ablation study conducted on EdgeSRGAN for four different steps. The last step corresponds to the final model. Overall, we observe -3.7\% LPIPS perceptual quality and +280\% inference speed. \scalebox{0.8}{\textbf{↑}}: higher is better, \scalebox{0.8}{\textbf{↓}}: lower is better.}
\label{tab:ablation}
\end{table}

\subsection{Application: Image Transmission for Mobile Robotics} 
\label{sec:use-case}

\begin{figure}[t]
	\centering
		\includegraphics[width=\textwidth]{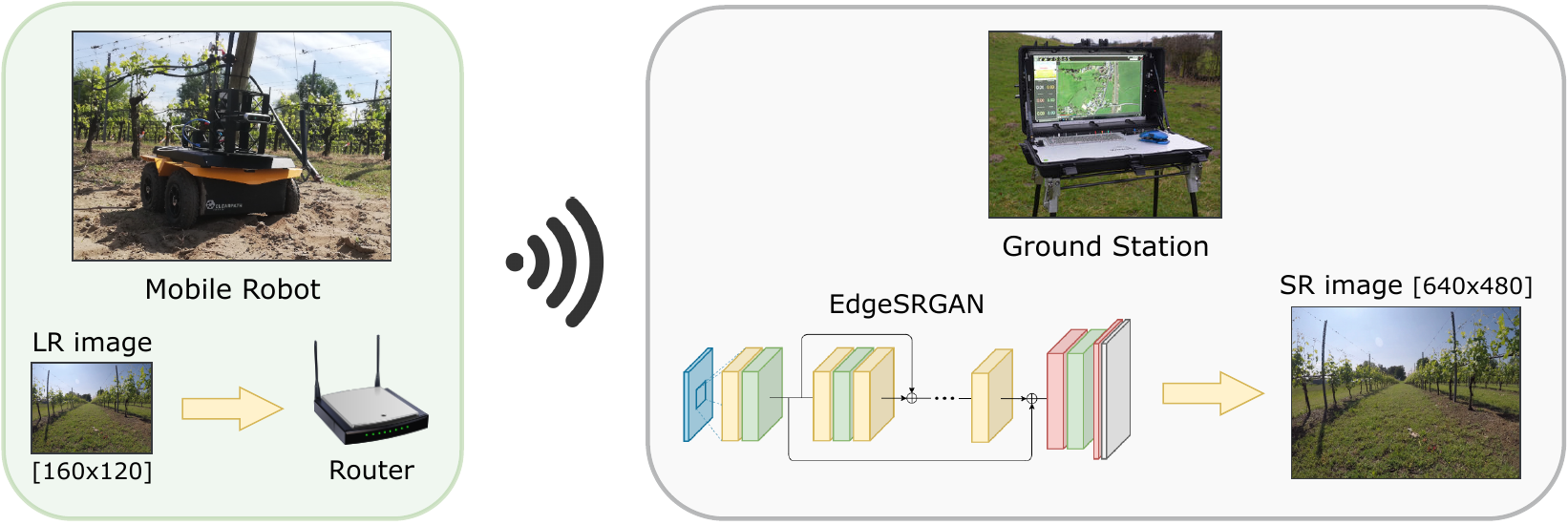}
	  \caption{Efficient image transmission system with EdgeSRGAN for mobile robotic applications in outdoor environments.}\label{fig:robot_scheme}
\end{figure}

Our real-time SISR can provide competitive advantages in a wide variety of practical engineering applications. In this section, we target a specific use case of mobile robotics, proposing our EdgeSRGAN system as an efficient deep learning-based solution for real-time image transmission. Indeed, robot remote control in unknown terrains needs a reliable transmission of visual data at a satisfying framerate, preserving robustness even in bandwidth-degraded conditions. This requirement is particularly relevant for high-speed platforms and UAVs. Dangerous or delicate tasks such as tunnel exploration, inspection, or open space missions all require an available visual stream for human supervision, regardless of the autonomy level of the platform.
In the last few years, the robotics community has focused on developing globally shared solutions for robot software and architectures and handling data communications between multiple platforms and devices.
ROS2 \cite{doi:10.1126/scirobotics.abm6074} is the standard operative system for robotic platforms. It is a middleware based on a Data Distribution System (DDS) protocol where application nodes communicate with each other through a topic with a publisher/subscriber mechanism. However, despite the most recent attempts to improve the reliability and efficiency of message and data packet communications between different nodes and platforms, heavier data transmission, such as image streaming, is not yet optimized and reliable.

The typical practical setting used for robot teleoperation and exploration in unknown environments is composed of a ground station and a rover connected to the same wireless network. As shown in Fig. \ref{fig:robot_scheme}, we adopted this ground station configuration to test the transmission of images through a ROS2 topic, as should be done in any robotic application to stream what the robot sees or to receive visual data and feed perception and control algorithms for autonomous navigation and mapping. For this experiment, we use both an Intel RealSense D435i camera\footnote{\url{https://www.intelrealsense.com/depth-camera-d435i/}} and a Logitech C920 webcam\footnote{\url{https://www.logitech.com/it-it/products/webcams/c920-pro-hd-webcam.960-001055.html}} mounted on a Clearpath Jackal robot\footnote{\url{https://clearpathrobotics.com/jackal-small-unmanned-ground-vehicle/}}, together with a Microhard BulletPlus\footnote{\url{https://www.microhardcorp.com/BulletPlus-NA2.php}} router for image transmission. The available image resolutions with RealSense cameras, the standard RGBD sensors for visual perception in robotics, are $(320\times240)$ and $(640\times480)$, whereas the framerate typically varies between 15 and 30 fps.

\begin{figure*}[t]
\begin{subfigure}[!htbp]{1.00\linewidth}
    \captionsetup{font=small}
    \centering
	\scriptsize
	\renewcommand{\g}{-0.7mm}
	\renewcommand{\tabcolsep}{1.8pt}
	\renewcommand{\arraystretch}{1}
	\resizebox{\linewidth}{!}{
	\newcommand{\name}{1_}
		\begin{tabular}{cccc}
		    \includegraphics[height=0.15\textwidth]{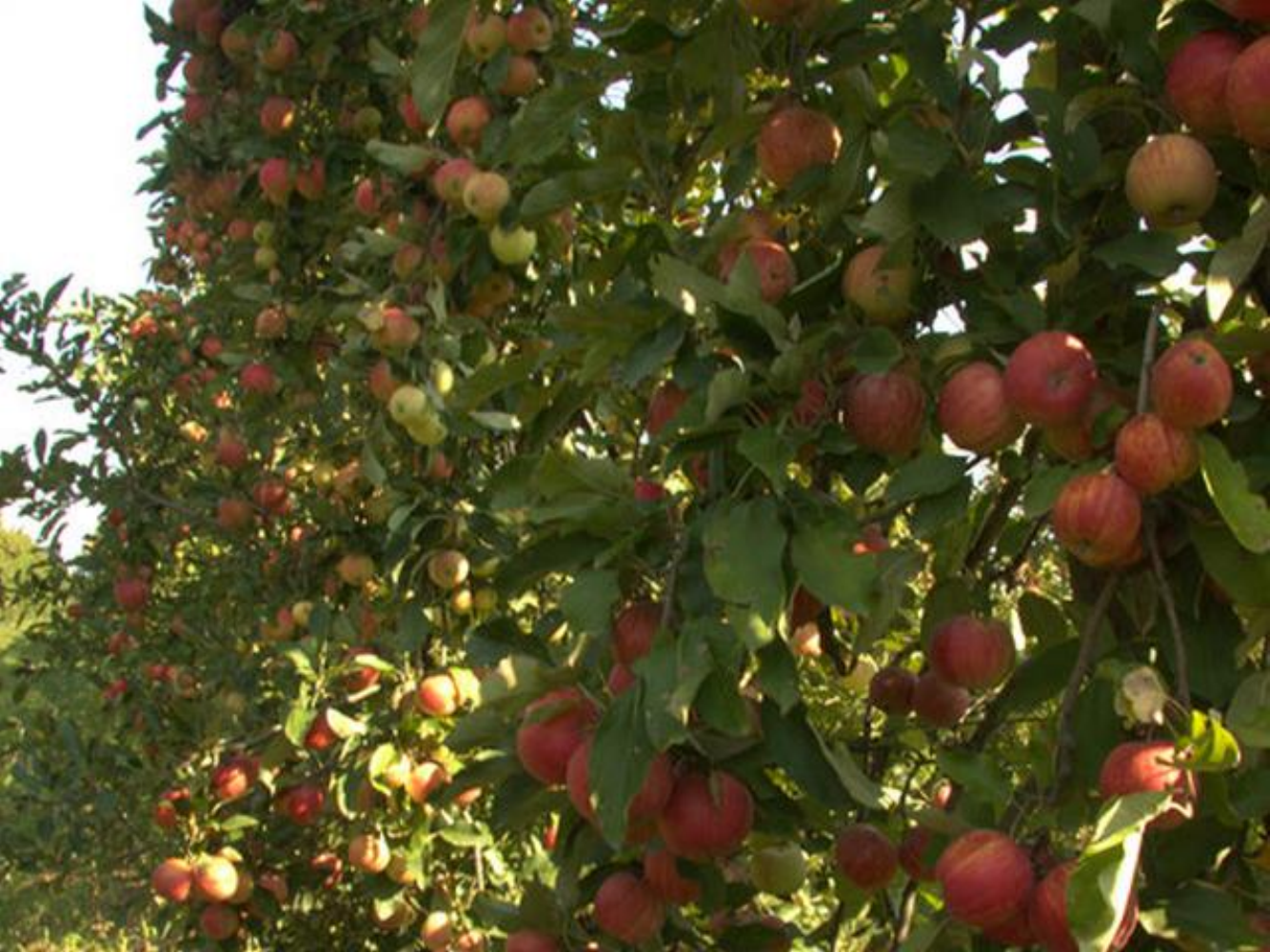} \hspace{\g} &
			\includegraphics[height=0.15\textwidth]{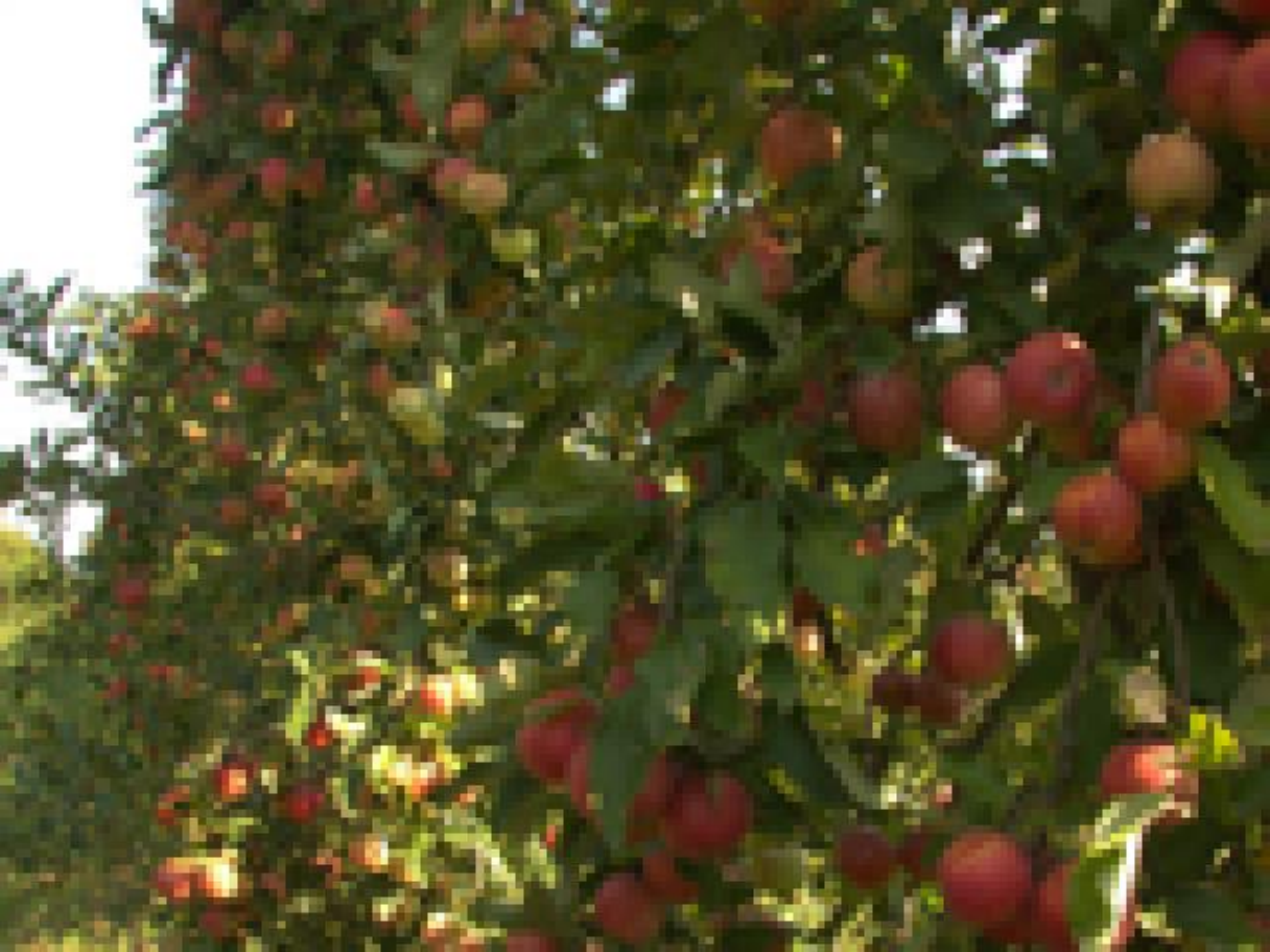} \hspace{\g} &
            \includegraphics[height=0.15\textwidth]{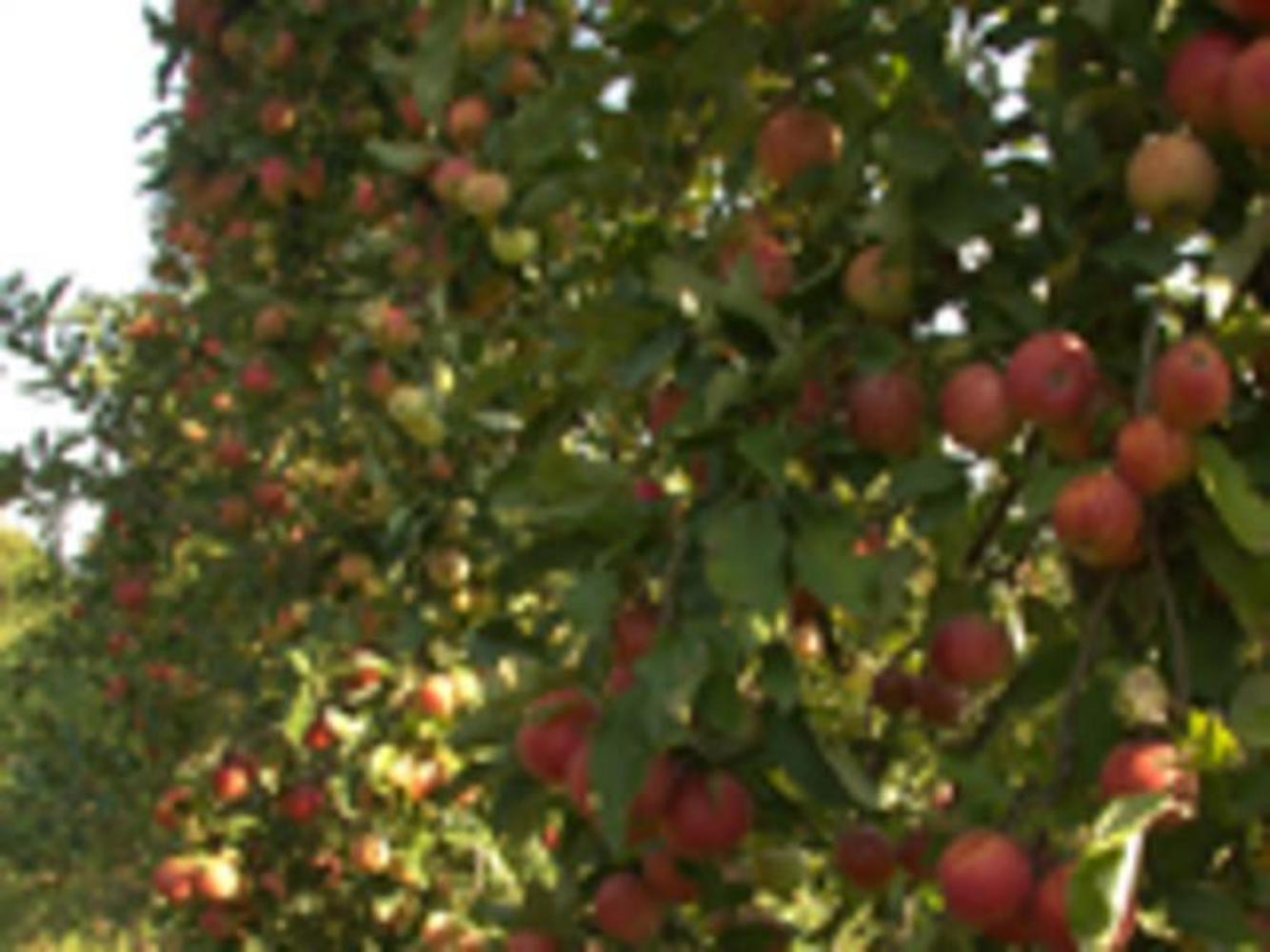} \hspace{\g} &
            \includegraphics[height=0.15\textwidth]{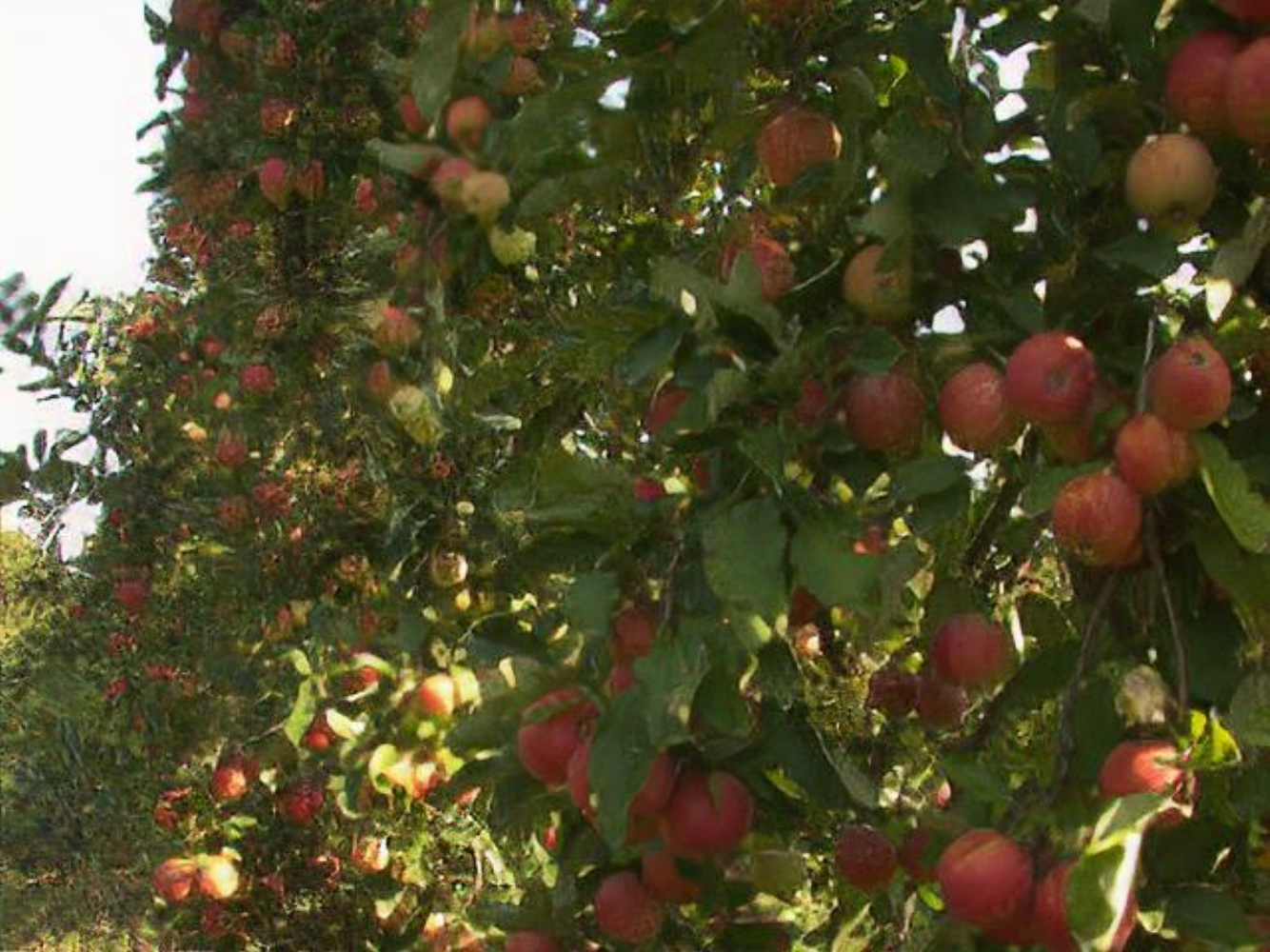} \hspace{\g}
			\\
			HR \hspace{\g} &
			LR \hspace{\g} &
            Bicubic \hspace{\g} &
            \textbf{EdgeSRGAN (ours)} \hspace{\g}
			\\
		\end{tabular}
	}
	\label{fig:real_apple}
\end{subfigure}
\begin{subfigure}[!htbp]{1.00\linewidth}
    \captionsetup{font=small}
    \centering
	\scriptsize
	\renewcommand{\g}{-0.7mm}
	\renewcommand{\tabcolsep}{1.8pt}
	\renewcommand{\arraystretch}{1}
	\resizebox{\linewidth}{!}{
	\newcommand{\name}{2_}
		\begin{tabular}{cccc}
		    \includegraphics[height=0.15\textwidth]{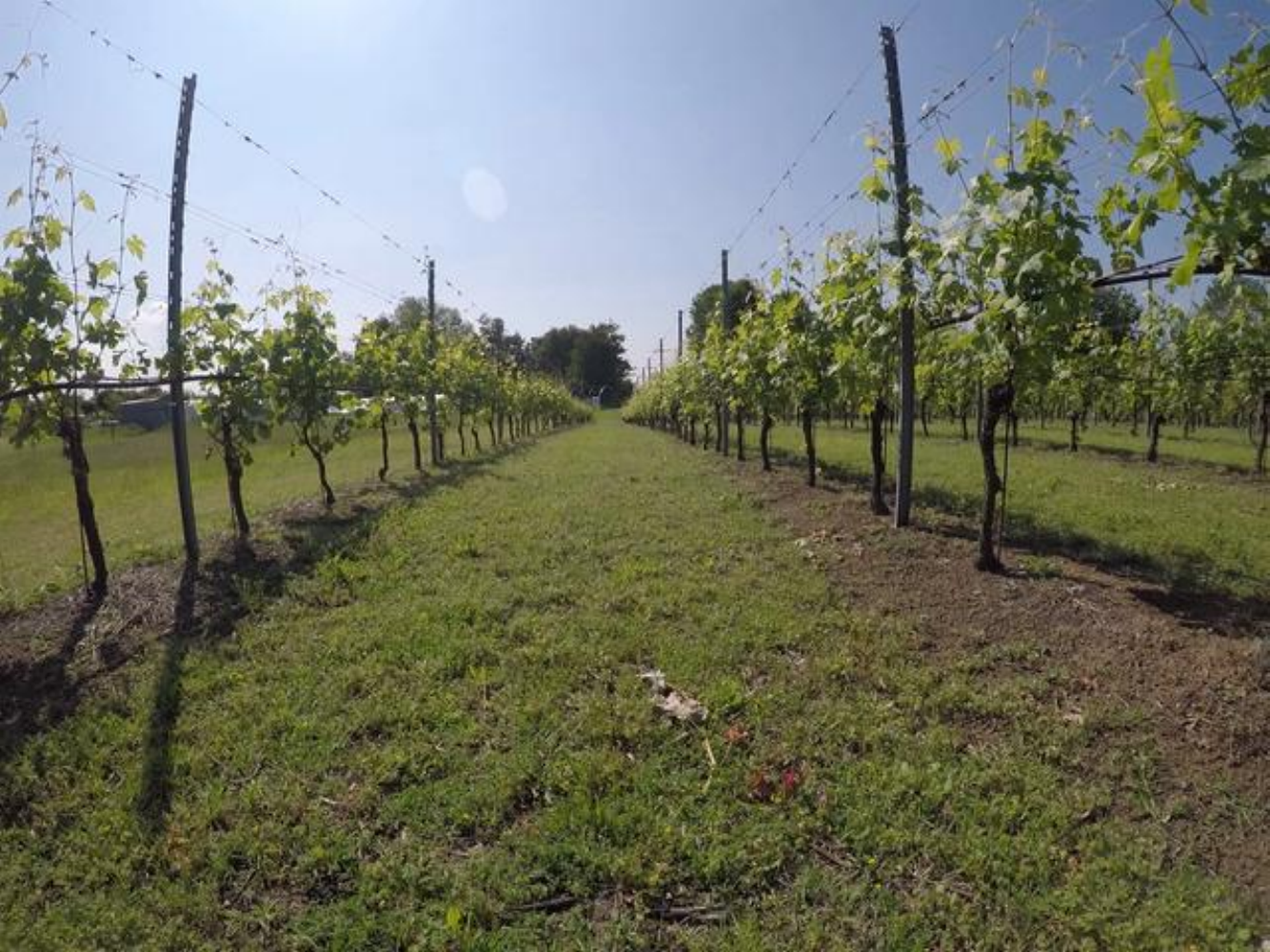} \hspace{\g} &
			\includegraphics[height=0.15\textwidth]{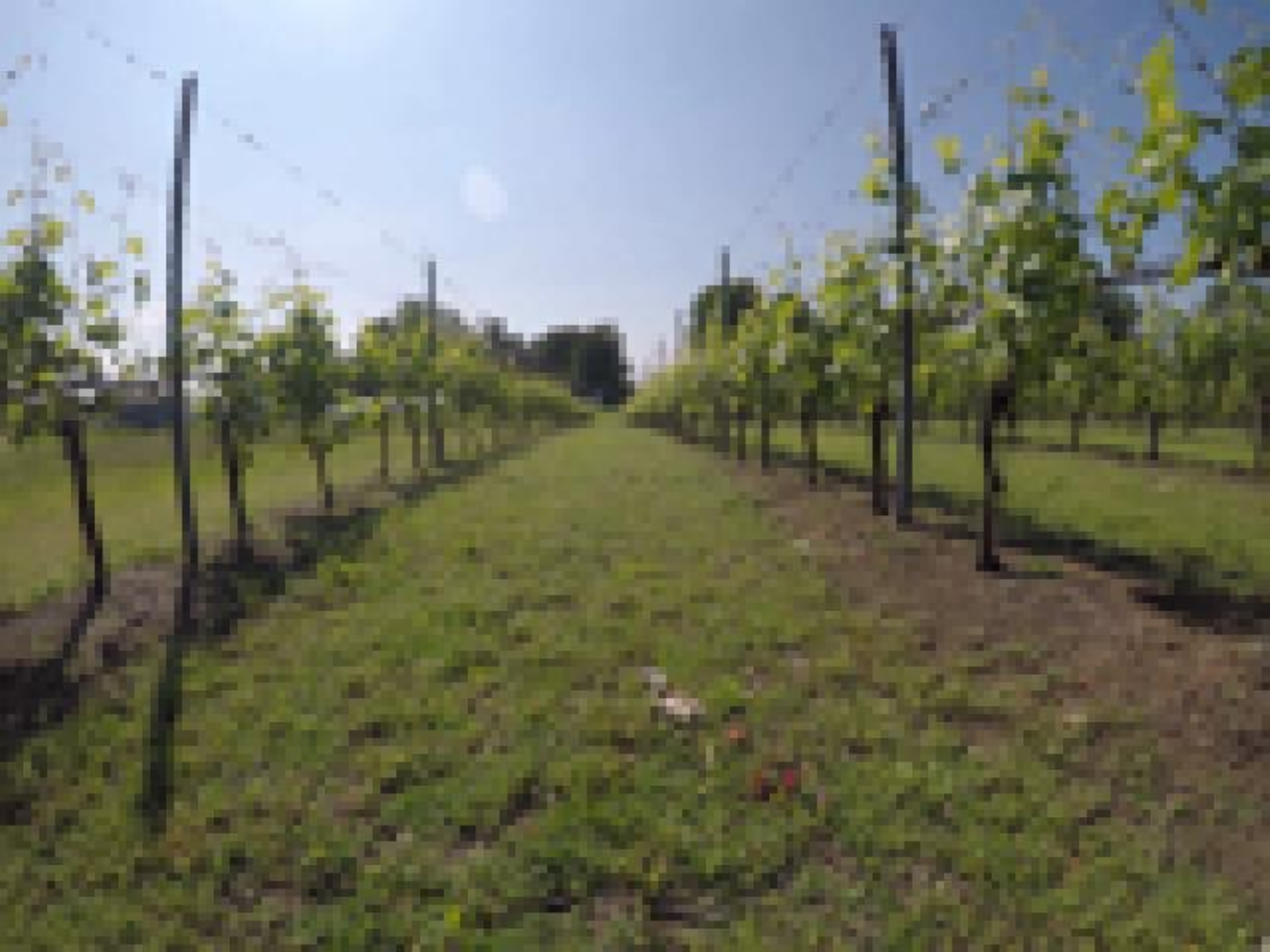} \hspace{\g} &
            \includegraphics[height=0.15\textwidth]{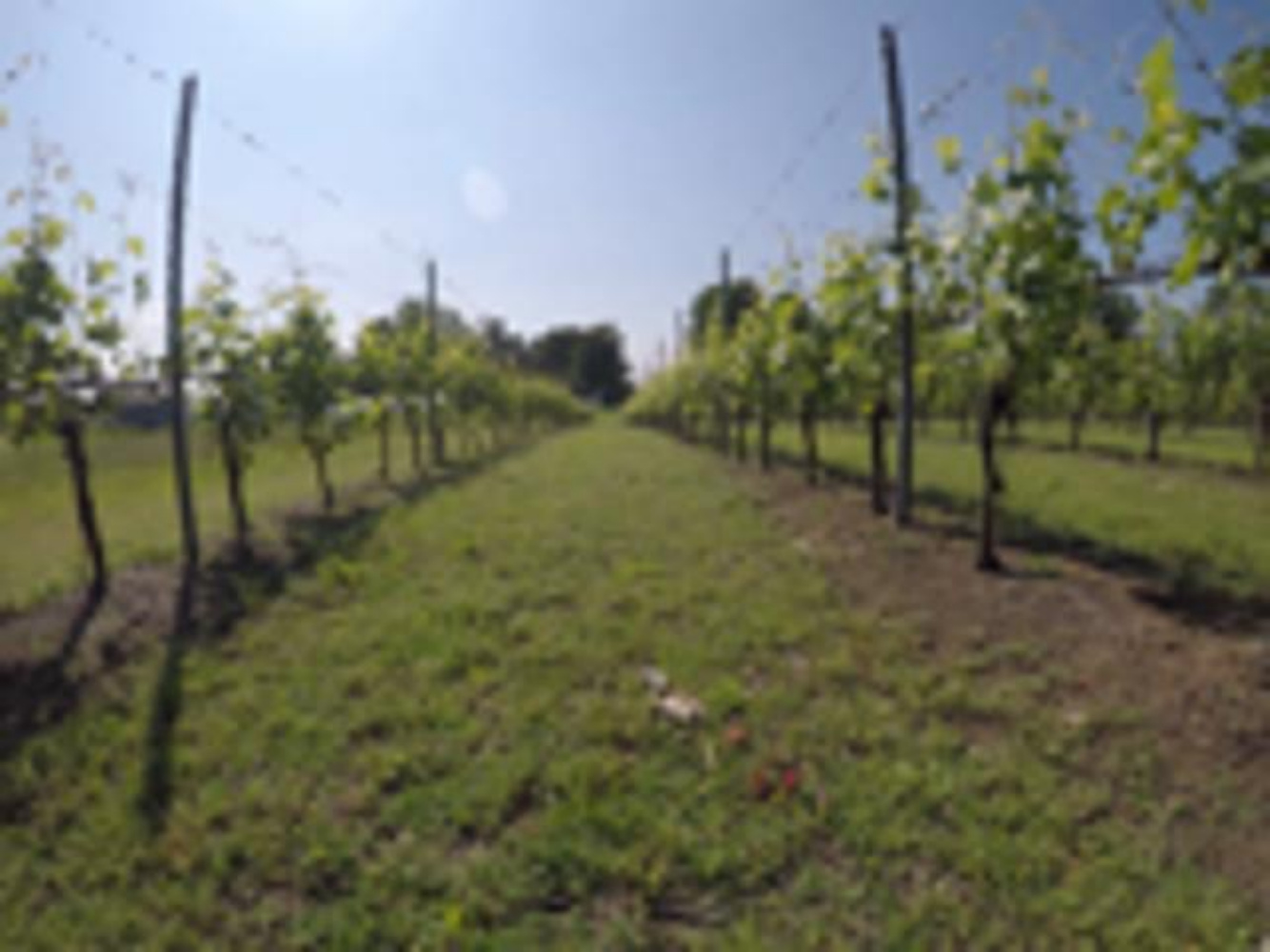} \hspace{\g} &
            \includegraphics[height=0.15\textwidth]{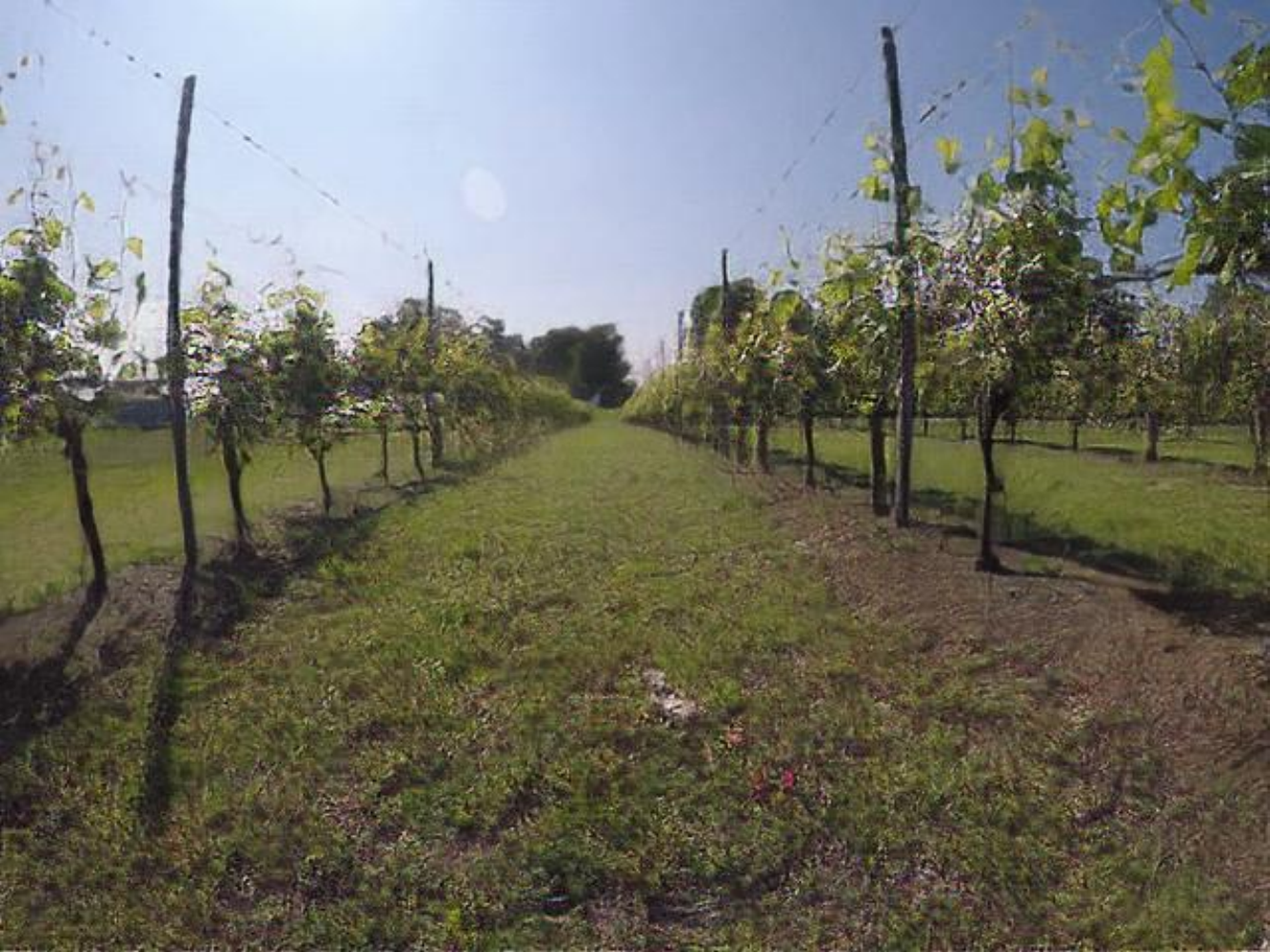} \hspace{\g}
			\\
			HR \hspace{\g} &
			LR \hspace{\g} &
            Bicubic \hspace{\g} &
            \textbf{EdgeSRGAN (ours)} \hspace{\g}
			\\
		\end{tabular}
	}
	\label{fig:real_row}
\end{subfigure}
\begin{subfigure}[!htbp]{1.00\linewidth}
    \captionsetup{font=small}
    \centering
	\scriptsize
	\renewcommand{\g}{-0.7mm}
	\renewcommand{\tabcolsep}{1.8pt}
	\renewcommand{\arraystretch}{1}
	\resizebox{\linewidth}{!}{
	\newcommand{\name}{3_}
		\begin{tabular}{cccc}
		    \includegraphics[height=0.15\textwidth]{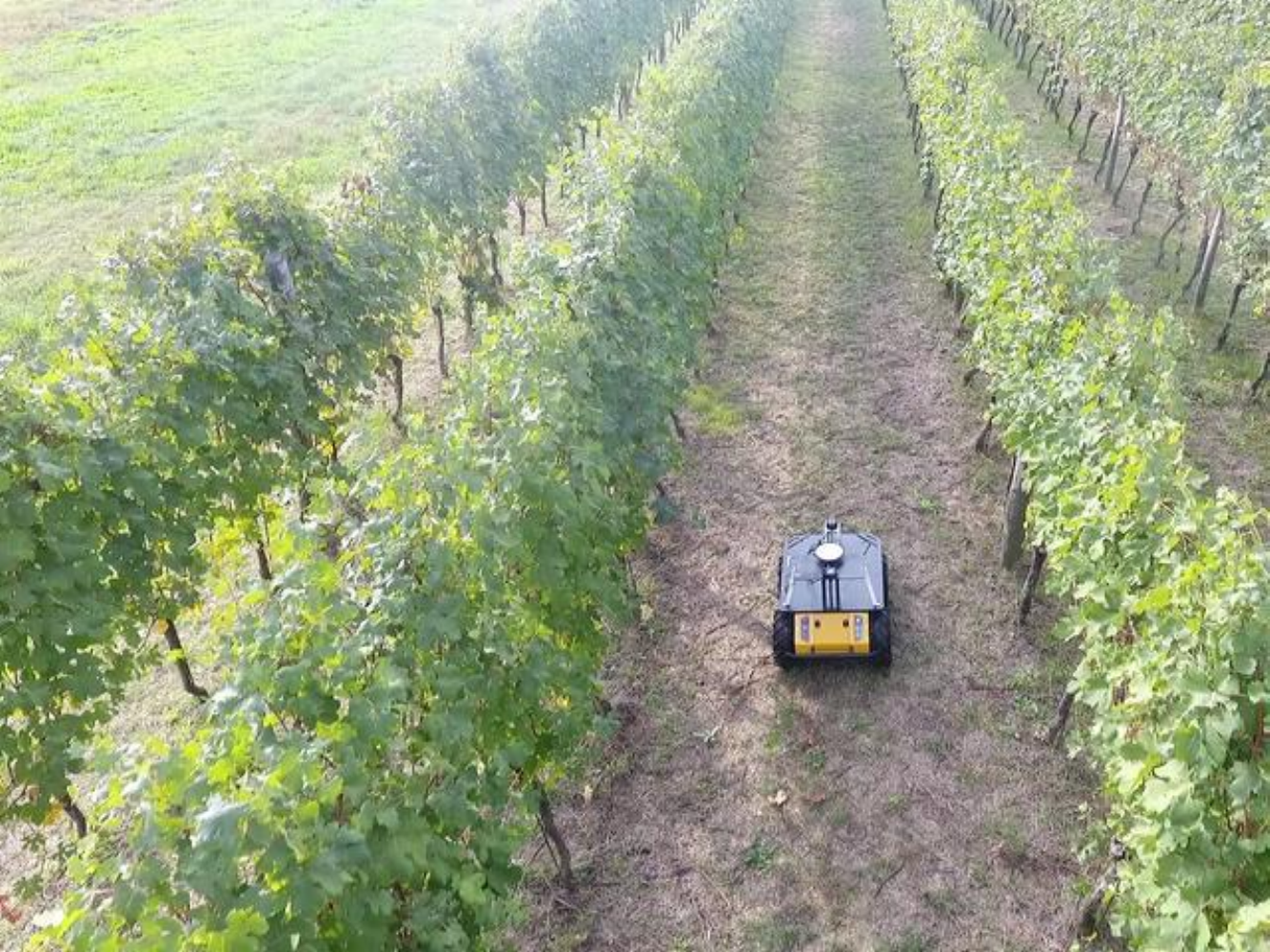} \hspace{\g} &
			\includegraphics[height=0.15\textwidth]{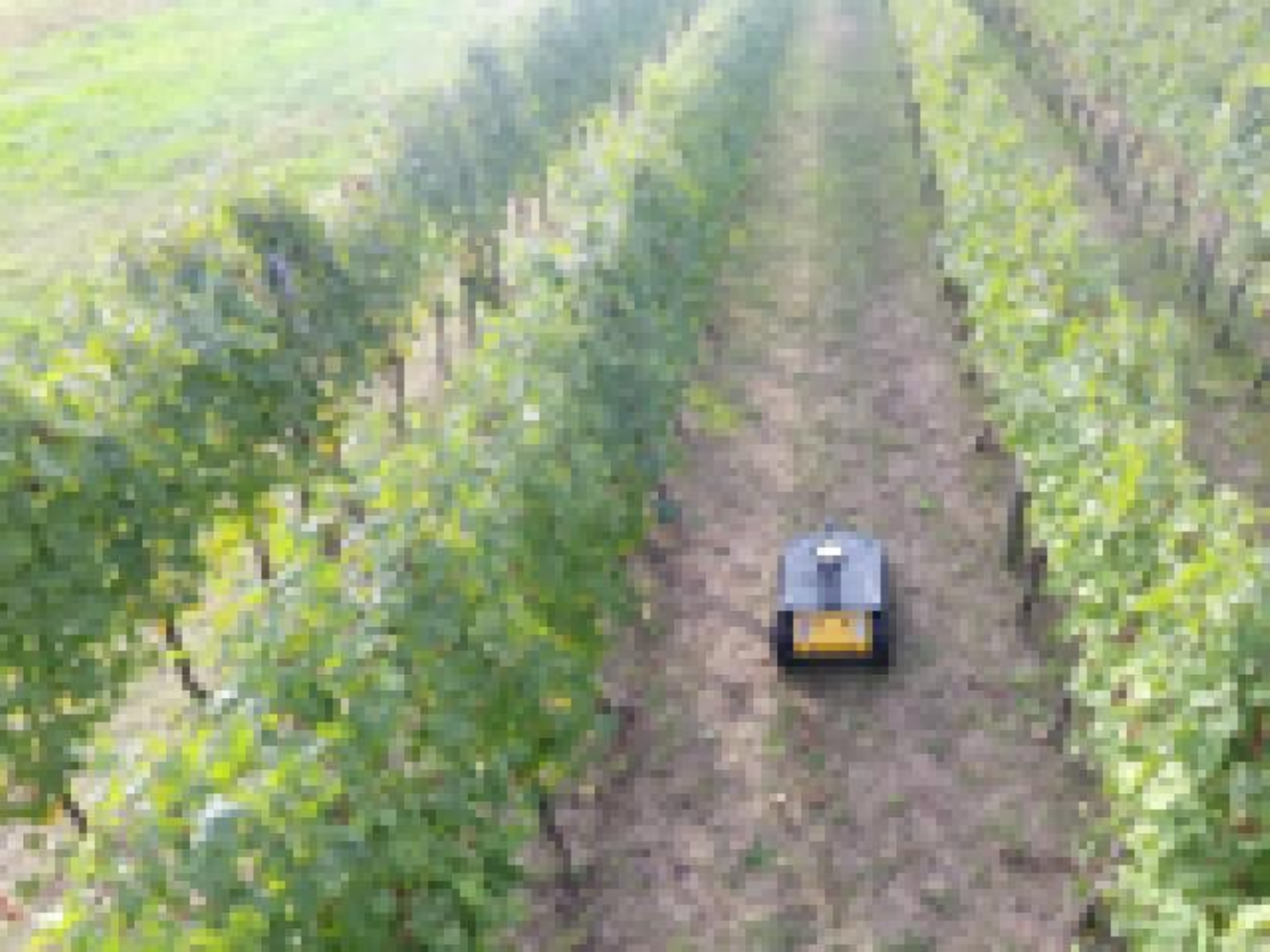} \hspace{\g} &
            \includegraphics[height=0.15\textwidth]{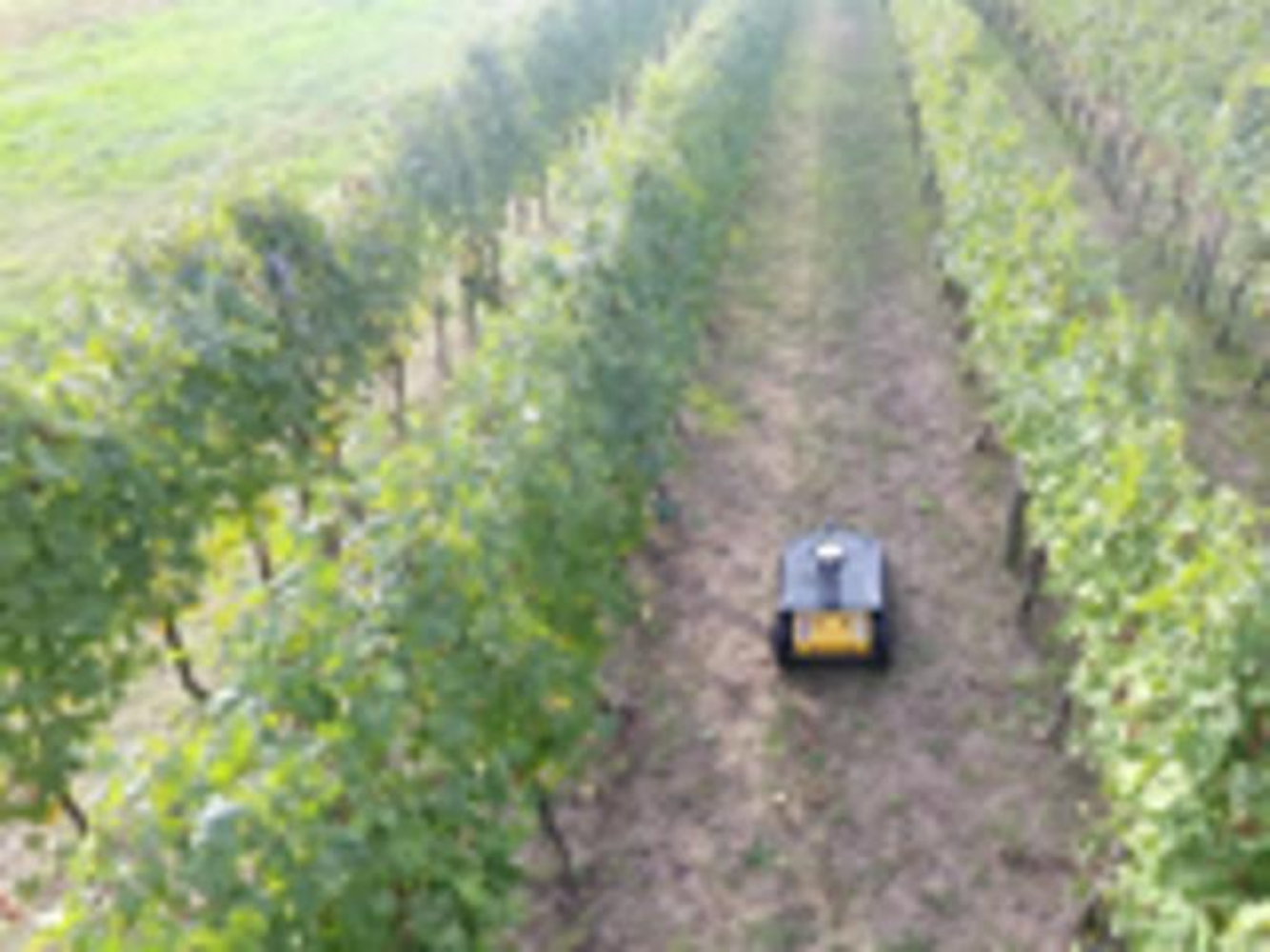} \hspace{\g} &
            \includegraphics[height=0.15\textwidth]{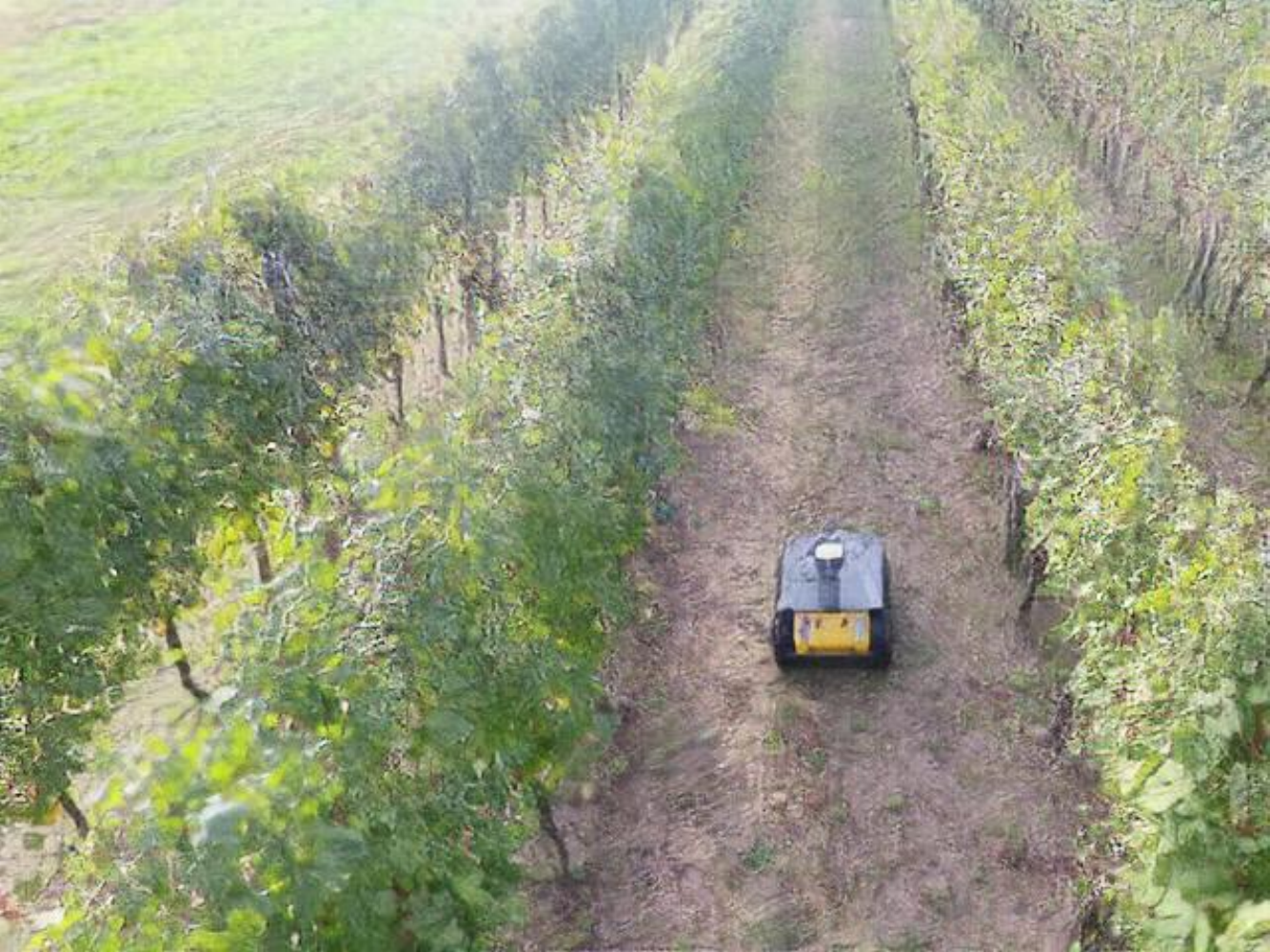} \hspace{\g}
			\\
			HR \hspace{\g} &
			LR \hspace{\g} &
            Bicubic \hspace{\g} &
            \textbf{EdgeSRGAN (ours)} \hspace{\g}
			\\
		\end{tabular}
	}
	\label{fig:real_drone}
\end{subfigure}
\begin{subfigure}[!htbp]{1.00\linewidth}
    \captionsetup{font=small}
    \centering
	\scriptsize
	\renewcommand{\g}{-0.7mm}
	\renewcommand{\tabcolsep}{1.8pt}
	\renewcommand{\arraystretch}{1}
	\resizebox{\linewidth}{!}{
	\newcommand{\name}{4_}
		\begin{tabular}{cccc}
		    \includegraphics[height=0.15\textwidth]{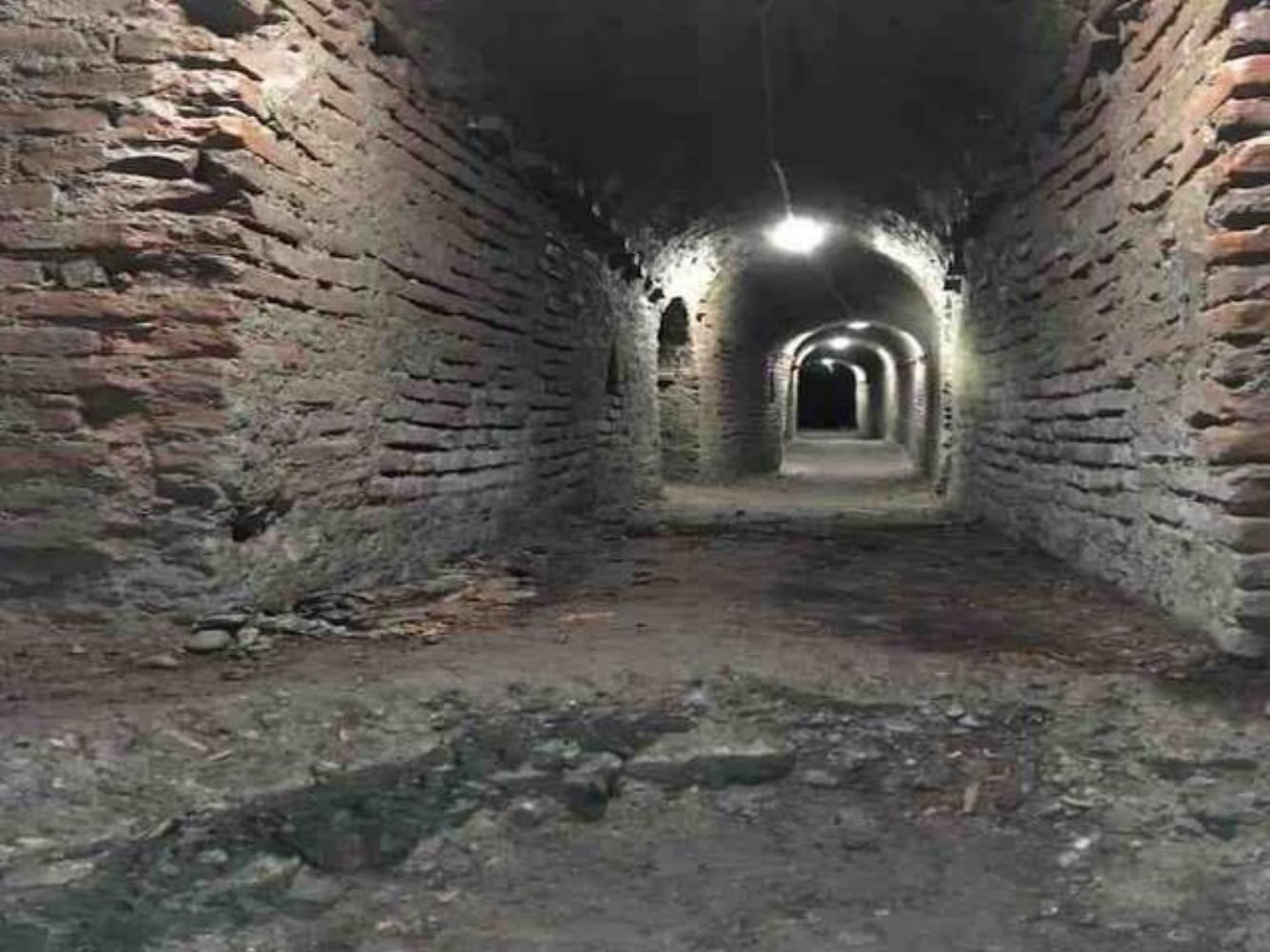} \hspace{\g} &
			\includegraphics[height=0.15\textwidth]{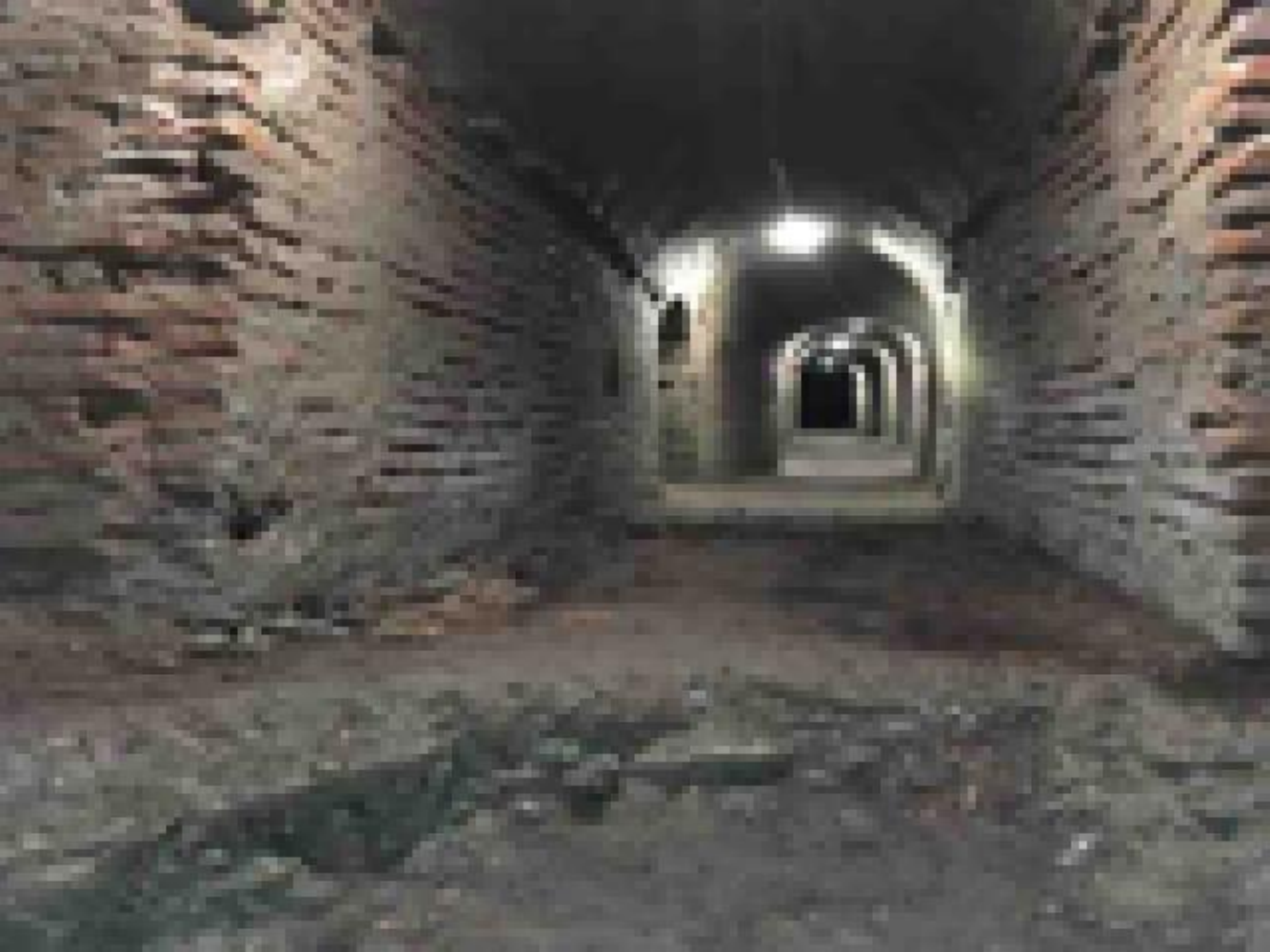} \hspace{\g} &
            \includegraphics[height=0.15\textwidth]{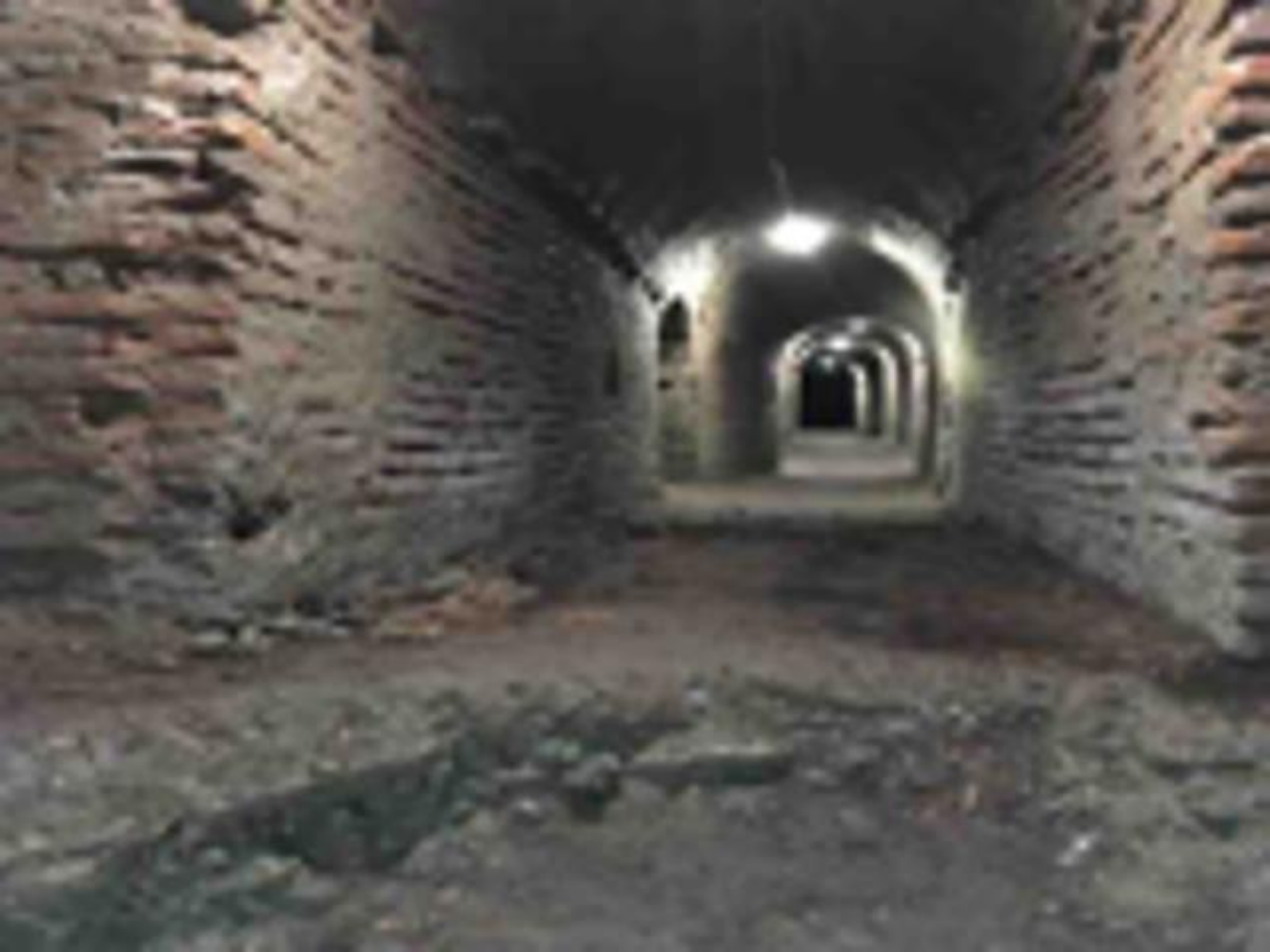} \hspace{\g} &
            \includegraphics[height=0.15\textwidth]{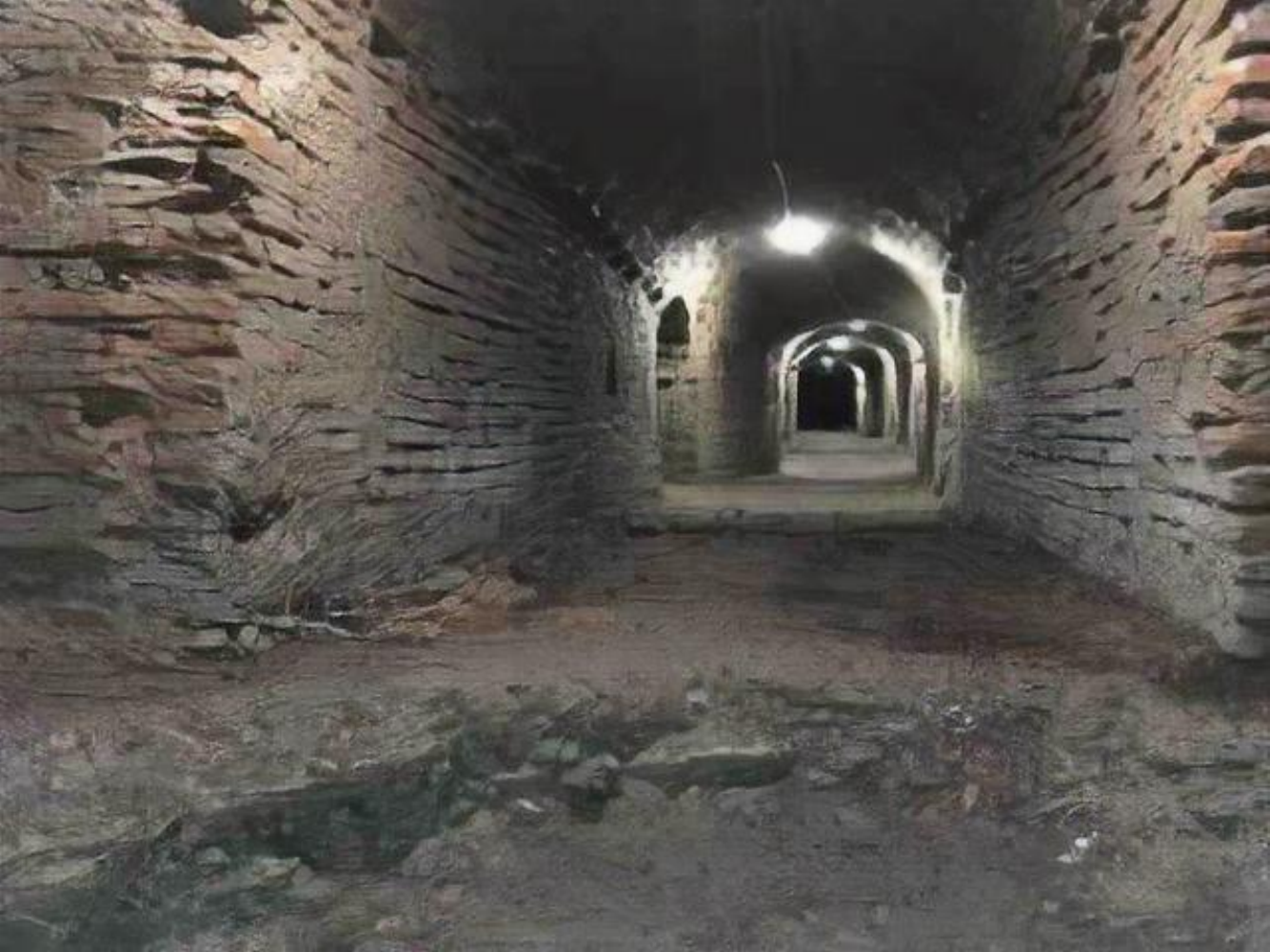} \hspace{\g}
			\\
			HR \hspace{\g} &
			LR \hspace{\g} &
            Bicubic \hspace{\g} &
            \textbf{EdgeSRGAN (ours)} \hspace{\g}
			\\
		\end{tabular}
	}
	\label{fig:real_tunnel}
\end{subfigure}
\vskip 8pt
\caption{Qualitative demonstration of applying EdgeSRGAN ($\times 4$) on real scenarios (zoom for more detail). From top to bottom: apple monitoring, navigation in vineyards, drone surveillance for autonomous rovers, and tunnel inspection.}
\label{fig:real_comparison}
\end{figure*}

Despite the absence of strong bandwidth limitations, transmission delays, or partial loss of packets, the maximum resolution and framerate allowed by ROS2 communication are extremely low: we find that at 30 fps, the maximum transmissible resolution for RGB is $(120\times120)$ with a bandwidth of 20 Mb/s while reducing the framerate to 5 fps the limit is $(320\times240)$. This strict trade-off between framerate and resolution hinders the high-speed motion of a robotic platform in a mission, increasing the risk of collision due to reduced scene supervision. Even selecting \textit{best effort} in the Quality of Service (QoS) settings, which manage the reception of packages through topics, the detected performances are always scarce.

Adopting our real-time Super-Resolution system ensures the timely arrival of RGB and depth images via ROS2. Thanks to the fast-inference performance of EdgeSRGAN, we can stream low-resolution images ($80\times60$) at a high framerate (30 fps) and receive a high-resolution output: $(320\times240)$ with a x4 image upsampling and $(640\times480)$ with a x8 upsampling, showing a clear improvement on standard performance. Our system allows the ground station to access the streaming data through a simple ROS topic. Hence, it provides multiple competitive advantages in robotic teleoperation and autonomous navigation: high-resolution images can be directly exploited by the human operator for remote control. Moreover, they can be used to feed computationally hungry algorithms like sensorimotor agents, visual-odometry, or visual-SLAM, which we may prefer to run on the ground station to save the constrained power resources of the robot and significantly boost the autonomy level of the mission. In Fig. \ref{fig:real_comparison}, we report a qualitative comparison to highlight the effectiveness of EdgeSRGAN for real-world robotic scenarios. In particular, we consider apple monitoring, navigation in vineyards, drone surveillance for autonomous rovers, and tunnel inspection. 

\begin{figure}[t]
    \centering
    \begin{subfigure}{0.49\linewidth}
        \centering
        \includegraphics[width=0.95\columnwidth, trim={19px 0 40px 0}, clip]{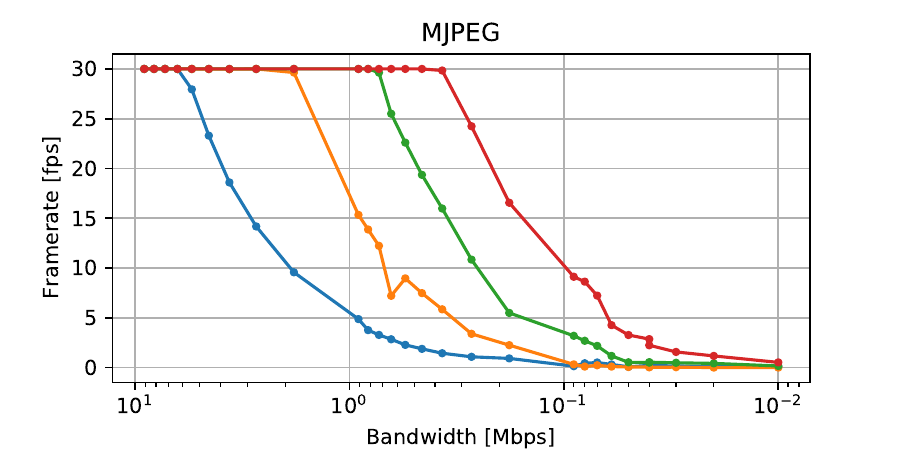}
    \end{subfigure}
    \begin{subfigure}{0.49\linewidth}
        \centering
        \includegraphics[width=0.95\columnwidth, trim={19px 0 40px 0}, clip]{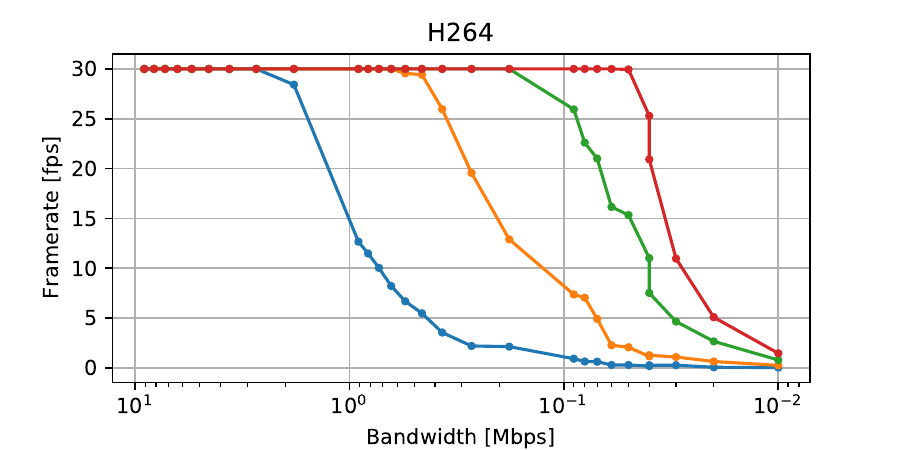}
    \end{subfigure}
    \begin{subfigure}{\linewidth}
        \centering
        \includegraphics[width=0.45\columnwidth, trim={10px 10px 10px 10px}, clip]{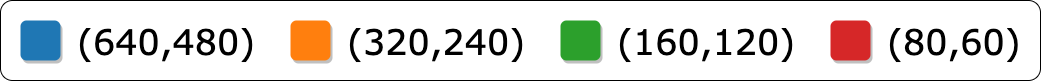}
    \end{subfigure}
	\caption{Framerate results vs. bandwidth for video transmission at different input resolutions with MJPEG and H264 compression. Bandwidth is in log scale.}\label{fig:band}
\end{figure}

We also test video transmission performance in a more general framework to reproduce all the potential bandwidth conditions. We use the well-known video streaming library GStreamer\footnote{\url{https://gstreamer.freedesktop.org/}} to transmit video samples changing the available bandwidth. We progressively reduce the bandwidth from 10 Mbps to 10 kbps using the Wondershaper library\footnote{\url{https://github.com/magnific0/wondershaper}} and measure the framerate at the receiver side. We use 10 seconds of the standard video sample \textit{smtpe} natively provided by GStreamer \textit{videotestsrc} video source at 30 fps, and we encode it for transmission using MJPEG and H264 video compression standards. The encoding is performed offline to ensure that all the available resources are reserved for transmission only. Indeed, most cameras provide hardware-encoded video sources without requiring software compression. To be consistent with the other experiments, we keep using $(640\times480)$ and $(320\times240)$ as high resolutions and $(160\times120)$ and $(80\times60)$ as low resolutions. Each experiment is performed 10 times to check the consistency in results. Fig. \ref{fig:band} presents the average framerate achieved with different bandwidths. Streaming video directly without any middleware, such as ROS2, ensures a higher transmission performance. However, as expected, streaming high-resolution images is impossible in the case of low bandwidth and the framerate quickly drops to very low values, resulting unsuitable for real-time applications.
On the other hand, lower resolutions can be streamed with minimal frame drop, even with lower available bandwidths. H264 compression shows the same behavior as MJPEG but shifts to lower bandwidths. Indeed, H264 is more sophisticated and efficient, as it uses temporal frame correlation in addition to spatial compression. In a practical application with a certain bandwidth constraint, a proper combination of a low-resolution video source and an SR model can be selected to meet the desired framerate requirements on the available platform (CPU or Edge TPU).
This mechanism can also be dynamically and automatically activated and deactivated depending on the current connectivity to avoid framerate drops and ensure a smooth image transmission.

\section{Conclusions and Future Works}\label{sec:concl}
In this paper, we proposed a novel Edge AI model for SISR exploiting the Generative Adversarial approach. Inspired by popular state-of-the-art solutions, we design EdgeSRGAN, which obtains comparable results, being an order of magnitude smaller in terms of the number of parameters. Our model is 3 times faster than SRGAN, 30 times faster than ESRGAN, and 50 times faster than SwinIR while retaining similar or even better LPIPS performance. To gain additional inference speed, we applied knowledge distillation to EdgeSRGAN and obtained an even smaller network (EdgeSRGAN-tiny) which gains an additional 4x speed with limited performance loss. Moreover, model quantization is used to optimize the model for execution on an Edge TPU. At the same time, network interpolation was implemented to allow potential users to balance the model output between pixel-wise fidelity and perceptual quality. Extensive experimentation on several datasets confirms the effectiveness of our model regarding both performance and latency. Finally, we considered the application of our solution for robot teleoperation, highlighting the validity and robustness of EdgeSRGAN in many practical scenarios in which the transmission bandwidth is limited. 
Future work may investigate the effect of additional optimization techniques, such as pruning \cite{li2016pruning} and neural architecture search \cite{pham2018efficient}. Moreover, developing optimized Edge AI versions of more recent architectures like transformers \cite{liang2021swinir} might bring advantages in tackling real-time SISR.

\section*{Acknowledgements}
This work has been developed with the contribution of the Politecnico di Torino Interdepartmental Center for Service Robotics PIC4SeR\footnote{\url{https://pic4ser.polito.it}} and SmartData@Polito\footnote{\url{https://smartdata.polito.it}}.
\bibliographystyle{abbrv}  
\bibliography{biblio}  
\clearpage
\end{document}